\documentclass[letterpaper,twocolumn,preprintnumbers,amsmath,amssymb,floatfix,nofootinbib]{revtex4-1}
\usepackage{graphicx} 
\usepackage{bm}
\usepackage{amsmath}
\usepackage{amsfonts} 
\usepackage{simplewick}
\usepackage{xcolor,multirow,longtable,url}

\begin{document}
\title{Physical features of strength of isoscalar pairing interaction determined by relation between double charge change and double pair transfer}
%
\author{J.~Terasaki}
\affiliation{Institute of Experimental and Applied Physics, Czech Technical University in Prague, Husova 240/5, 100$\,$00 Prague 1, Czech Republic}
\date{Mar.~6,~2020}
 
\begin{abstract}
A new method has been proposed to determine the strength of the isoscalar proton-neutron pairing interaction applicable to many nuclei. The principle is the equivalence between the double charge change and the double transfer of like-particle pair, and a constraint is derived to the effective interactions used in approximations. This method was applied to the quasiparticle random-phase approximation for determining that interaction strength. In this paper, detail of this method is explained thoroughly, and applications are made to nuclei of several instances of the double-$\beta$ decays. The systematics of the strengths determined for those nuclei is understood in terms of a  midshell effect. The effect of the new interaction strength is examined in two examples of the Gamow-Teller strength function with comparisons with the experimental data. The nuclear matrix elements of the neutrinoless double-$\beta$ decay are also calculated. 
\end{abstract}

\maketitle

\section{Introduction \label{sec:introduction}}
The proton-neutron (pn) pairing correlations are an interesting and important subject in nuclear physics. The physically unique point is the possibility of the unlike-particle pair. The history of studies of this subject has a variety of ideas and approaches. 
These correlations are usually discussed with the classification of the isospin $T=0$ (isoscalar) or 1 (isovector). The former is spin triplet, and the latter is spin singlet. This classification is useful because the isospin is an approximate good quantum number for nuclei with a good accuracy. From this isospin symmetry of nuclei, the nuclear isospin properties obey the linear algebra under the rotation of the nucleus in the isospin space. The neutron-neutron, proton-proton, and pn $T$$=$$1$ pairing interactions are thought to be three components of a vector in the isospin space specified by the $z$-component of the isospin  $T_z=1, -1$ and 0, respectively. According to this idea, the strength of the isovector pn pairing interaction is expected to be close to those of the like-particle (lp) pairing interactions. On the other hand, one does not have a guideline as this algebraic idea for determining the strength of the $T$$=$$0$, necessarily pn, pairing interaction. This is one of the reasons why the isoscalar pairing correlations have been an issue in nuclear physics. The question is how those correlations manifest themselves, and for the practitioners, how the interaction strength can be determined for applications and predictions. The spacial overlap of the density distributions of the single-particles are essential for creating the pair by a short-range interaction. Thus, the pn pairing correlations are anticipated to manifest themselves in nuclei with the same proton number $Z$ and neutron number $N$ more strongly than the others. 

The method is established to obtain the experimental lp pairing gap from the systematics of the masses using the three-point or four-point formula \cite{Boh69}. The strength of the pairing interaction is usually determined so as to reproduce this experimental gaps by the calculation. If there is no gap, the system is not in the pair condensate. In other words, if there is a finite pairing gap, that entire value implies that the system is in the pair condensate. Such a physical quantity is most appropriate for determining the strength of the pairing interaction by the fitting. 
It seems, therefore, reasonable to consider an analysis for the isoscalar pairing gaps analogous to that established method. This approach has been investigated in Ref.~\cite{Mac00a}. If the topmost proton and neutron in the single-particle level scheme of odd-odd $N$$=$$Z$ nuclei form a pair, the masses of the odd-odd nuclei  should be systematically close to the line obtained by the interpolation of the masses of even-even nuclei as a function of the mass number $A$. The authors of Ref.~\cite{Mac00a} investigated this possibility for the $N$$=$$Z$ nuclei with $A=8$$-$$60$ (the ground states have $T$ = 0) and clarified that the  masses of the odd-odd nuclei are deviated from the mass systematics of the even-even nuclei by 2.5$-$7.5 MeV. They concluded that there is no evidence for an isoscalar pair condensate in the $N$$=$$Z$ nuclei. In another paper \cite{Mac00b}, they discuss the spectrum of the pn pairing vibrational states around $^{56}$Ni and conclude that the isoscalar channel does not have appreciable collectivity. They state in that paper that the strength of the isoscalar pairing interaction is much smaller than that of the isovector one. 

The authors of Ref.~\cite{Sat01} investigated the mass systematics in terms of the Wigner energy. That is an extra binding of the $N$$=$$Z$ nuclei compared to the smoothly interpolated energy from the energies of the neighboring nuclei in the $N$$+$$Z$$=$constant line. They determined the strengths of the isoscalar pairing interaction to reproduce the trend of the experimental Wigner energy for $A$ = 24$-$52 by their extended Hartree-Fock-Bogoliubov (HFB)  calculation and showed that $^{24}$Mg and $^{48}$Cr, as representatives, have finite isoscalar pairing gaps. In fact, their isoscalar pairing gaps are much larger than the isovector ones. They also performed the isospin-cranking calculation and reproduced  the excitation energies of the $T$$=$$2$ excited states using their interaction strengths. The isoscalar-to-isovector ratio of their pairing interaction strengths is $\approx$1.65 at the beginning of the \textit{sd} shell and  $\approx$1.4 in the $f_{7/2}$ subshell with a smooth decrease. 
The above two studies seem to imply that the understanding of the isoscalar pairing correlations is totally different depending on whether one refers to the mass systematics along the $N$$=$$Z$ line or the line perpendicular to it. This conflict is also one of the reasons for the difficulty in understanding the isoscalar pairing correlations. 

The authors of Ref.~\cite{Pov98} investigated the Wigner energy using the shell model. They performed the shell-model calculation with the Kuo-Brown interaction KB3 for the $N$$=$$Z$ nuclei ($A$ = 48 and 50) and calculated the Wigner energy. They also calculated the isoscalar and isovector pairing energies using the pairing interaction derived from their interaction according to the method developed in Ref.~\cite{Duf96}. That is a method to derive generally the separable interactions from the interactions for the shell model. The authors of the shell-model paper  showed that both the $T$$=$$0$ and 1 pairing energies are much smaller than the Wigner energy and concluded that they find no link between the Wigner energy and the dominant pairing terms of the nuclear interaction. Thus, the physical origin of the Wigner energy is a problem. The isoscalar-to-isovector ratio of their pairing interaction strengths is 1.56.  

The authors of Ref.~\cite{Lan97} also investigated the pn pairing correlations using the Shell Model
Monte Carlo (SMMC) method for the $N$$=$$Z$ and neighboring nuclei in the $pf$ shell using the KB3 interaction. They calculated the expectation values of the number of the pairs of nucleons coupled to the angular momentum $J$ = 0 (isovector) and 1 (isoscalar). The results of their calculation show a few unique features.  Firstly, in the iron isotopes with large neutron excess, the isoscalar pn correlations dominate over the isovector ones in terms of the pair number. This result implies that the closeness of the Fermi surfaces of the protons and neutrons is not necessary for the isoscalar pairing correlation contrary to the simple picture mentioned above. Secondly, it was shown  for $A$ = 48$-$60 ($N$ = $Z$) that the isoscalar pairing correlations increase as $A$ increases. The HFB calculations, e.g., \cite{Sat01}, show that the heavier $N$$=$$Z$ nuclei, the smaller the isoscalar pairing gaps. The reason is thought to be that the effect of the spin-orbit interaction increases for the nuclei involving the single-particles with larger angular momentum, e.g., \cite{Pov98}. The single-particles in the spin-orbit partner can form a $T$$=$$0$ pair, but the spin-orbit splitting is disadvantageous compared to the degenerated orbits forming the $T$$=$$1$ pair. Thirdly, the pair numbers of the mean-field wave functions were also calculated. Those results have the same tendency on the above two features, and the values of the pair number are often close to those of the SMMC method. This result is also unique because the pairing correlations are conceptually not compatible with the independent-particle picture. However, the pair number can be finite as the expectation  values of any other operators conserving the particle number are so. 

The density-functional approach (the mean-field approximation and its extensions) constructs the interactions or energy-density functional step by step by adding the components for extending the many-body correlations; an example is the pairing interactions. This extension is made by referring to the experimental data, as long as the relevant data are available. On the other hand, the shell model uses the effective interactions with which it is expected to obtain every many-body effect correctly when the dynamical equation is solved. Thus, it is a reasonable approach to determine the isoscalar pairing interaction from the shell-model interactions, as an example was mentioned above.  
The authors of Ref.~\cite{Ber10} considered all the shell-model interaction matrix elements with total spin and isospin couplings $(J, T)$ = $(0, 1)$, $(1, 0)$ and made a least-squares fit to each set using the contact interaction and harmonic oscillator orbitals. The interactions referred to were the USDB Hamiltonian fitted to \textit{sd}-shell nuclei and the GX1A Hamiltonian fitted to \textit{fp}-shell nuclei. The isoscalar-to-isovector ratio of their pairing interaction strengths is 1.65 for the \textit{sd} shell and 1.63 for \textit{fp} shell. They performed HFB calculations using that contact interaction as the pairing interaction and the Woods-Saxon and the spin-orbit potentials. The correlation energy was calculated which was defined as the energy in the absence of the pair condensate subtracted by the total HFB energy. In the calculation for $^{48}$Cr with the full \textit{fp} space, it was shown that the correlation energy with $(0,1)$ is larger than that with $(1,0)$. According to their interpretation, the ground state should exhibit ordinary pairing. They also showed that the isoscalar pairing can be dominant in much heavier nuclei of $A\sim 130$$-$140. This is the new point of their study. They argue that the spin-orbit field is ineffective ``at controlling the single-particle spectrum'' in the limit of large nuclear size because the spin-orbit field has a surface nature. An approach based on the shell-model interaction has also been applied for the nuclear matter \cite{Gar01}. 

A method independent of other methods to determine the strength of the isoscalar pairing interaction was used in Ref.~\cite{Tan14}. The authors of this paper used a simple equation which relates the interaction strength, the scattering length of the proton and neutron, and the effective range related to the cutoff energy or momentum. This method is an application of the method \cite{Ber91, Esb97} originally used for the neutron-neutron pairing. 
They obtained the isoscalar-to-isovector ratio of the contact density-dependent interaction of 1.9 with the cutoff energy of 20 MeV. They performed the three-body calculations (valence proton and neutron and a core) for odd-odd $N$$=$$Z$ nuclei with $A = 7$$-$29 and discussed the spectra of the lowest two states with $(J^\pi, T)=(0^+,1)$ and $(1^+,0)$. The systematics of the experimental spectra sometimes shows the inversion of the two levels, and the correct orderings were reproduced by their calculations. They state that the spin-orbit splitting prevents the strong isoscalar pairing interaction and makes the ground states of $^{34}$Cl and $^{42}$Sc to have $J ^\pi$ = 0$^+$. According to this study, the low-lying spectra of those nuclei are strongly affected by the isoscalar pairing correlations. 

It is also a reasonable approach to exploit physical quantities strongly reflecting on the isoscalar pairing correlations because the isoscalar pairing gap is not established. 
Calculations were performed of the Gamow-Teller (GT) strength function by using the pn random-phase approximation (RPA) \cite{Bai13}. The comparison with the experimental data of $^{56}$Ni shows that the isoscalar interaction improves the calculated GT strength function, however pinning down of the best strength of that interaction is difficult because the perfect reproduction of the energy dependence of the GT strength is difficult. 
The authors of Ref.~\cite{Bai14} showed that the behavior of the cumulative sum of the GT strength of $^{42}$Ca  with respect to the excitation energy is reproduced very well by the isoscalar pairing interaction, of which the strength is larger than that of the isovector pairing interaction by a factor of 1.05. 

Measurements of cross sections of $(p,^3$He) and ($^3$He,$p$) reactions were made for several $N$$=$$Z$ $sd$-shell nuclei at forward angles \cite{ayy17}. They obtained the ratios of the $J^\pi$$=$$0^+$ cross section to   the $J^\pi$$=$$1^+$ one and compared them to those calculated without the pairing interactions. It was found that the experimental ratio is appreciably lower than that of the no-pairing calculation for $^{24}$Mg($^3$He,$p$)$^{26}$K, and they concluded that $^{24}$Mg has strong isoscalar correlations. The slight lowering in the same analysis was also found for $^{40}$Ca($p$,$^3$He)$^{38}$K. The results of shell-model calculations are also shown; the tendency is similar to the experimental data. To my knowledge, these data are not yet used for determining the isoscalar pairing interactions. The deuteron transfer with $J^\pi=1^+$ is expected to show enhancement, if the nucleus has the strong isoscalar pairing correlations. This speculation is inferred in analogy with the pairing vibrations of the like-particles studied for the Pb region by the two-particle transfer reactions \cite{Bes71}. 

The authors of Ref.~\cite{Fea06} showed measured $B(M1)$ of $^{46,48}$Ti and compared them with shell-model results with and without the isoscalar pairing interaction. Without the isoscalar pairing interaction the calculated $M1$ strengths in a low-energy region $\leq$ 8 MeV are not sufficient compared to the experimental data, and with that interaction the calculation results are closer to the data. It is shown that the difference is in the spin component of the $M1$ transition strength. 

The strength of the isoscalar pairing interaction has also been studied in relation to the double-$\beta$ decay. The nuclei most intensively considered are the candidates of the neutrinoless double-$\beta$ (0$\nu\beta\beta$) decay, which is a key point for determining the neutrino mass scale; see e.g., \cite{Eng17}. The strength of the isoscalar pairing interaction is crucial to the pn quasiparticle RPA (pnQRPA) approach to the nuclear matrix element of this decay because it is known that this matrix element is sensitive to that interaction \cite{Vog86,Civ87}. Two important parameters are not given a priori. One is the strength of the isoscalar pairing interaction, and another is the effective axial-vector current coupling $g_A$ for the double-$\beta$ decays. The latter parameter is the strength of the GT component of the weak interaction. The half-lives to the two-neutrino double-$\beta$ ($2\nu\beta\beta$) decay of all the candidate nuclei used for the  $0\nu\beta\beta$ experiments are known experimentally \cite{Bar19}. These data can be used by fitting for removing an uncertainty of the parameters \cite{Rod03}, however, still the combination of $g_A$ and the isoscalar pairing strength has an uncertainty. 
The authors of Ref.~\cite{Sim09} performed the pnQRPA calculations for obtaining the $0\nu\beta\beta$ nuclear matrix elements using G-matrixes of NN interactions with the pairing interaction proportional to the G-matrix interaction with a modifying factor. They calculated two $0\nu\beta\beta$ nuclear matrix elements for each of eight decay instances with  $g_A$ = 1.254 (the value for a free nucleon\footnote{The recent value is  $\simeq$1.27.}) and 1.0 (a typical effective value, e.g., \cite{Bro85}) and determined the modifying factor for the pairing interaction so as to reproduce the measured half-lives of the $2\nu\beta\beta$ decays. 
There are variations in the scheme to determine the parameters; see, e.g., Refs.~\cite{Vog86,Suh05,Sim18}. 
There are many other studies on the pn pairing correlations; see, e.g., Ref.~\cite{Goo79} for early studies. 

In this paper, I investigate a method independent of others for determining the strength of the isoscalar pairing interaction. The starting point of this method is an identity implying that the effective interactions have a relation.  This idea was created in the study of the $0\nu\beta\beta$ decay \cite{Ter16}. My motivation is to remove an uncertainty in the application of  the QRPA to the calculation of the nuclear matrix element of that $\beta\beta$ decay. All nuclei treated by the experiments for finding the $0\nu\beta\beta$ decay are $N$$\neq$$Z$ nuclei, and the experimental data of those nuclei for discussing the isoscalar pairing correlations are much less abundant than  those of the $N$$=$$Z$ nuclei. A method is necessary to determine the strength of the isoscalar pairing interaction in this situation. My method discussed in this paper is a general method satisfying this requirement. This paper is organized as follows: Sec.~\ref{sec:new_idea} presents the basic idea of the method together with the mathematical preparations.  The applications to some of the candidate nuclei of the $0\nu\beta\beta$ decay are shown in Sec.~\ref{sec:application}. The physical features of the obtained strengths of the isoscalar pairing interactions are discussed in Sec.~\ref{sec:properties_strengths}. Tests of the obtained strength are shown in terms of the GT strength functions in Sec.~\ref{sec:comparison_str_exp}, and the calculated nuclear matrix elements of the $0\nu\beta\beta$ decays are shown in Sec.~\ref{sec:NME_0vbb}. Section \ref{sec:summary} is the summary. 

\section{\label{sec:new_idea} Formulation}

\subsection{\label{sec:basic_idea} Principle to determine strength of isoscalar pairing interaction }
Let me consider the calculation of the transition matrix element
\begin{eqnarray}
M^{(\textrm{\scriptsize{dcc}})} = \langle F | T^{(\textrm{\scriptsize{dcc}})} | I \rangle , 
\label{eq:Mdcc}
\end{eqnarray}
where $|I\rangle$ and $|F\rangle$ are the ground states, obtained by the QRPA, of even-even nuclei with different proton number $Z$ and neutron number $N$. It is assumed that $|I\rangle$ has $(Z,N)$, and $|F\rangle$ has $(Z+2,N-2)$. The transition operator is a double-charge-change operator 
\begin{eqnarray}
T^{(\textrm{\scriptsize{dcc}})} = \sum_{pp^\prime nn^\prime} \langle pp^\prime | V(\bm{r}) | nn^\prime \rangle 
c^\dagger_{p^\prime} c_{n^\prime} c^\dagger_p c_n .
\end{eqnarray}
Symbols $p, p^\prime$ and $n,n^\prime$ denote the proton and neutron states, respectively, and $c^\dagger_i$ and $c_i$ ($i$: single-particle state) are the creation and annihilation operators, respectively. $V(\bm{r})$ is a two-body potential ($\bm{r}$ is the relative-position vector of two nucleons) including the operator changing two neutrons to two protons. My idea is not affected by the $\bm{r}$ dependence of $V$. Thus, $V(\bm{r})$ is arbitrary in this section, as long as the two-body matrix elements do not vanish.  

The pnRPA \cite{Hal67} is a useful method for describing the charge-change phenomena. I use the quasiparticle version of pnQRPA. 
By this method the creation operators $O^\dagger_{B^\textrm{\tiny{pn}}_I}$ of single-charge-changed states $|B^\textrm{\scriptsize{pn}}_I\rangle$ are obtained;  
\begin{eqnarray}
O^\dagger_{B^\textrm{\tiny{pn}}_I}|I\rangle = |B^\textrm{\scriptsize{pn}}_I\rangle.  
\end{eqnarray}
$O^\dagger_{B^\textrm{\tiny{pn}}_I}$ is expressed using creation and annihilation operators of quasiparticle $\{ a^\dagger_\mu, a_\nu \}$ as a linear combination of $a^\dagger_p a^\dagger_n$'s and their Hermite conjugate operators, and the higher-order components are ignored. Note that $a^\dagger_p a_n$ is of the higher order than $a^\dagger_p a^\dagger_n$ and $a_p a_n$ in the pnQRPA. These creation operators satisfy 
\begin{eqnarray}
&& [O_{B^\textrm{\tiny{pn}}_I}, 
O^\dagger_{B^{\prime\textrm{\tiny{pn}}}_I} ] 
= \delta_{ B^\textrm{\tiny{pn}}_I,B^{\prime\textrm{\tiny{pn}}}_I }, \nonumber\\ 
&& [O_{B^\textrm{\tiny{pn}}_I}, O_{B^{\prime\textrm{\tiny{pn}}}_I} ] = 0,  \label{eq:boson_comm}
\end{eqnarray}
in the pnQRPA order. 
The transformation between 
$\{O^\dagger_{B^\textrm{\tiny{pn}}_I}, O_{B^\textrm{\tiny{pn}}_I}\}$ and $\{a^\dagger_p a^\dagger_n, a_n a_p \}$ is unitary (with a special metric), thus,  $c^\dagger_p c_n$ can be written as a linear combination of $O^\dagger_{B^\textrm{\tiny{pn}}_I}$ and $O_{B^\textrm{\tiny{pn}}_I}$ in the pnQRPA. 
Therefore, it holds that 
\begin{eqnarray}
\sum_{B^\textrm{\tiny{pn}}_I}|B^\textrm{\scriptsize{pn}}_I\rangle\langle B^\textrm{\scriptsize{pn}}_I| c^\dagger_p c_n|I\rangle = c^\dagger_p c_n |I\rangle,
\end{eqnarray}
again in the pnQRPA order.
Thus, Eq.~(\ref{eq:Mdcc}) can be written as   
\begin{eqnarray}
M^{(\textrm{\scriptsize{dcc}})} &=& \sum_{pp^\prime nn^\prime} \langle pp^\prime | V(\bm{r}) | nn^\prime \rangle 
\sum_{ B^\textrm{\tiny{pn}}_F }
\sum_{ B^\textrm{\tiny{pn}}_I }
\langle F | c^\dagger_{p^\prime} c_{n^\prime} | B^\textrm{\scriptsize{pn}}_F \rangle
\nonumber\\
&&\times \langle B^\textrm{\scriptsize{pn}}_F | B^\textrm{\scriptsize{pn}}_I \rangle 
\langle B^\textrm{\scriptsize{pn}}_I | c^\dagger_p c_n | I \rangle , \label{eq:Mdcc_dcc_path}
\end{eqnarray}
where $|B^\textrm{\scriptsize{pn}}_F\rangle$  is defined in the same way as $|B^\textrm{\scriptsize{pn}}_I\rangle$ but for $|F\rangle$. 

Another way of calculating $M^{(\textrm{\scriptsize{dcc}})}$ is to use the lpQRPA. The mathematical properties of this approximation theory are analogous to the pnQRPA. The only but important difference is that the two-quasiparticle creation operators of the like-particles $\{a^\dagger_\mu a^\dagger_{\nu}\}$ (both $\mu$ and $\nu$ are protons or neutrons) and their Hermite conjugate operators are used as the building blocks to construct the creation operators of the lpQRPA  states $|B^\mathrm{lp}_I\rangle$ and $|B^\mathrm{lp}_F\rangle$; $|B^\mathrm{lp}_I\rangle$ ($|B^\mathrm{lp}_F\rangle$) is obtained on the basis of $|I\rangle$ ($|F\rangle$).  These states are approximation of the excited states keeping the proton and neutron number and the states with the  proton or neutron numbers different by two from that of the ground state. If the HFB ground state has the pair condensate of like-particles, these two types of states are mixed in the lpQRPA states. 
Analogously to the application of the pnQRPA, one obtains another expression of Eq.~(\ref{eq:Mdcc})   
\begin{eqnarray}
M^{(\textrm{\scriptsize{dcc})}} &=& \sum_{pp^\prime nn^\prime} \langle pp^\prime | V(\bm{r}) | nn^\prime \rangle 
\sum_{ B^\textrm{\tiny{lp}}_F }
\sum_{ B^\textrm{\tiny{lp}}_I }
\langle F | c^\dagger_{p^\prime} c^\dagger_p | B^\textrm{\scriptsize{lp}}_F \rangle 
\nonumber \\
&&\times \langle B^\textrm{\scriptsize{lp}}_F | B^\textrm{\scriptsize{lp}}_I \rangle 
\langle B^\textrm{\scriptsize{lp}}_I |  c_n c_{n^\prime}| I \rangle . \label{eq:Mdcc_tpt_path}
\end{eqnarray}

In the QRPA, the equality of Eqs.~(\ref{eq:Mdcc_dcc_path}) and (\ref{eq:Mdcc_tpt_path}) is not an equation satisfied for arbitrary interactions because the many-body correlations taken into account are different for the two QRPA methods. 
This is evident by considering the pn pairing interaction. The lpQRPA-Hamiltonian matrix does not depend on this interaction, as long as the HFB ground state is not a pn-pair condensate. On the other hand, the pnQRPA-Hamiltonian matrix depends on that interaction. 
If exact nuclear wave functions are available, the equality of those two expressions would be guaranteed for any interactions. 
The key point of my new idea is to use this equality as a constraint to the effective interactions for  the QRPA. If the equality is satisfied for the effective interactions  used in the calculation, that calculation has no problem in terms of the theoretical consistency. The interaction suitable to adjust for satisfying the equality is the isoscalar pairing interaction, because it's strength is difficult to determine as clearly as the strength of the lp pairing interaction as reviewed in Sec.~\ref{sec:introduction}. 

For the isovector pn pairing interaction, a possible method to determine it's strength is to assume the isospin invariance of the isovector pairing interaction; see Sec.~\ref{sec:introduction}. Although the strengths of the proton-proton and neutron-neutron pairing interactions are not identical, it is a possible approximation to use the average value of the two strengths as the strength of the isovector pn pairing interaction. 
This prescription is used in my calculations. 
Other interactions are assumed to be established. Therefore, the proposed equality plays a role to determine the strength of the isoscalar pairing interaction.
\vfill

\subsection{\label{sec:examination} QRPA states}
The QRPA ground state is defined as the vacuum to the QRPA ``phonon";  
\begin{eqnarray}
&& O_{B^\textrm{\tiny{pn}}_I}|I\rangle = 0, \nonumber\\
&& O_{B^\textrm{\tiny{lp}}_I}|I\rangle = 0. \label{eq:QRPAvac_condition}
\end{eqnarray}
The pnQRPA and lpQRPA correlations are physically different type of correlations.  Thus, when both correlations are used in discussion, it is reasonable to consider that $|I\rangle$ is conceptually similar to the product states of the pnQRPA and  lpQRPA ground states (the explicit equation is shown below). 
Let $|I_{\textrm{\scriptsize{pn}}}\rangle$ and $|I_{\textrm{\scriptsize{lp}}}\rangle$ be the pnQRPA and lpQRPA ground states, respectively. 
The basic question is if, e.g., $\langle B^{\textrm{\scriptsize{pn}}}_I |c^\dagger_p c_n |I\rangle$ can be approximated by 
$\langle I_{\textrm{\scriptsize{pn}}}|O_{B^\textrm{\tiny{pn}}_I} c^\dagger_p c_n |I_{\textrm{\scriptsize{pn}}}\rangle$; this is the transition-density matrix element usually calculated by the pnQRPA. Below, the equations for investigating this approximation are presented. Those are equations discussed previously in Ref.~\cite{Ter16} except for Appendix, and this discussion is included here for self-containment. 

For answering that question, it is necessary to treat the explicit expressions of the pnQRPA ground state (see e.g., Ref.~\cite{Bal69})  
\begin{eqnarray}
&&|I_{\textrm{\scriptsize{pn}}}\rangle = \frac{1}{\mathcal{N}_{\mathrm{pn}I}} \prod_{K\pi} \exp [v^{K\pi}_{\mathrm{pn}I}] |i\rangle, \label{eq:pnQRPAvac} \\
[10pt]
&& \mathcal{N}_{\mathrm{pn}I}^{\,2} = 
\prod_{K\pi} \langle i | \exp [v^{K\pi\dagger}_{\mathrm{pn}I}]
\exp [v^{K\pi}_{\mathrm{pn}I}] | i\rangle, \label {eq:pnQRPAvac_norm} \\
[10pt]
&& v^{K\pi}_{\mathrm{pn}I} = 
\sum_{\mu\nu\mu^\prime\nu^\prime} C^{\textrm{\scriptsize{pn}}I,K\pi}_{\mu\nu,-\mu^\prime -\nu^\prime} a^{i\dagger}_\mu a^{i\dagger}_\nu a^{i\dagger}_{-\mu^\prime} a^{i\dagger}_{-\nu^\prime}, \label{eq:pnQRPAvac_v} \\
&&(\mu \textrm{ and } {-\mu^\prime}:\textrm{proton};\, \nu \textrm{ and } {-\nu^\prime}:\textrm{neutron}), \nonumber
\end{eqnarray}
where $|i\rangle$ is the HFB ground state, which is the vacuum for the quasiparticle;
\begin{eqnarray}
a^i_\mu |i\rangle = 0.
\end{eqnarray}
For the quasiparticle basis $\{a_\mu^{f\dagger}, a_\mu^f\}$ associated with the HFB state of the final nucleus $|f\rangle$, I have 
\begin{eqnarray}
a^f_\mu |f\rangle = 0,
\end{eqnarray}
and the equations analogous to Eqs.~(\ref{eq:pnQRPAvac})$-$(\ref{eq:pnQRPAvac_v}) can be written on the basis of the final state.  
Two quantum numbers are introduced; $K$ is the $z$-component of the nuclear angular momentum, and $\pi$ is the parity. Axially-symmetric deformed nuclei, and also spherical ones, are treated in my numerical calculations below. The ground state has $(K\pi)=(0+)$. Due to the time-reversal symmetry of the ground state, only $K\geq 0$ are treated explicitly in the numerical calculations. 
According to these nuclear symmetries, the quasiparticle states also have the quantum numbers of the $z$-component of the angular momentum $j^z_\mu$ and parity $\pi_\mu$. When $j^z_\mu$ of state $\mu$ is inverted, a label $-\mu$ is used. 
The coefficient $C^{\textrm{\scriptsize{pn}}I,K\pi}_{\mu\nu,-\mu^\prime -\nu^\prime}$ is called correlation coefficient \cite{Ull72} and obtained from the pnQRPA solutions. The four quasiparticle states $\{\mu,\nu,\mu^\prime,\nu^\prime\}$ in Eq.~(\ref{eq:pnQRPAvac_v}) are limited to those with 
\begin{eqnarray}
&&j^z_\mu + j^z_\nu = j^z_{\mu^\prime} + j^z_{\nu^\prime} = K, \label{eq:K_2qp}\\[5pt]
&& \pi_\mu \pi_\nu = \pi_{\mu^\prime} \pi_{\nu^\prime} = \pi. \label{eq:pi_2qp}
\end{eqnarray}
It has been confirmed \cite{Ter13} that the operators of the QRPA order ($\propto$ $a^{i\dagger}_\mu a^{i\dagger}_\nu, a^i_\nu a^i_\mu, a^{f\dagger}_\mu a^{f\dagger}_\nu$, or $a^f_\nu a^f_\mu$) with different $(K\pi)$ commute with each other with a good accuracy in the calculation of the overlap $\langle B^{ K\pi }_{\mathrm{lp}F} | B^{ K\pi }_{\mathrm{lp}I} \rangle$. Thus, the order of $\exp[v^{K\pi}_{ \mathrm{pn}I }]$'s in the product  in Eq.~(\ref{eq:pnQRPAvac}) is arbitrary. 
The creation operator of the pnQRPA state can be set to 
\begin{eqnarray}
&&O^\dagger_{B^{K\pi}_{\mathrm{pn}I}} = 
\sum_{\mu\nu} ( X^{\textrm{\scriptsize{pn}}I,K\pi}_{B,\mu\nu} 
a^{i\dagger}_\mu a^{i\dagger}_\nu - Y^{\textrm{\scriptsize{pn}}I,K\pi}_{B,{-\mu}{-\nu}} a^i_{-\nu} a^i_{-\mu}) , \label{eq:Opndagger} \\
&& (\mu:\textrm{proton}, \nu:\textrm{neutron}). \nonumber
\end{eqnarray}
In this equation, the label $B$ distinguishes the pnQRPA states having the $K$ and $\pi$,  and 
the pairs of $\mu\nu$ satisfy $j^z_\mu + j^z_\nu = K$ and $\pi_\mu\pi_\nu=\pi$. 
The forward amplitude $X^{\textrm{\scriptsize{pn}}I,K\pi}_{B,\mu\nu}$ and backward amplitude $Y^{\textrm{\scriptsize{pn}}I,K\pi}_{B,{-\mu}{-\nu}}$ are obtained by solving the pnQRPA equation. The two-quasiparticle pairs with specified $K$ and $\pi$  in Eq.~(\ref{eq:pnQRPAvac_v}) can be expressed by linear combinations of $O^\dagger_{B^{K\pi}_{\mathrm{pn}I}}$ and it's Hermite conjugates.  By applying the boson-like commutation relations (\ref{eq:boson_comm}) to the first one of Eq.~(\ref{eq:QRPAvac_condition}) it follows that 
\begin{eqnarray}
C^{\textrm{\scriptsize{pn}}I,K\pi}_{\mu\nu,{-\mu^\prime}{-\nu^\prime}} &=& 
\frac{1}{1+\delta_{K0}}\sum_B Y^{\textrm{\scriptsize{pn}}I,K\pi\ast}_{B,{-\mu^\prime} {-\nu^\prime}} \nonumber \\
&&\times \left(\frac{1}{ X^{\textrm{\scriptsize{pn}}I,K\pi\ast} }\right)_{B,\mu\nu}, \label{eq:correlation_coeff_pn}
\end{eqnarray}
where $1/X^{\textrm{\scriptsize{pn}}I,K\pi\ast}$ denotes the inverse matrix of which the row and column indexes are $B$ and $\mu\nu$, respectively. 

The corresponding equations of the lpQRPA are derived analogously;
\begin{eqnarray}
&&|I_{\textrm{\scriptsize{lp}}}\rangle = \frac{1}{\mathcal{N}_{\mathrm{lp}I}} \prod_{K\pi} \exp [v^{K\pi}_{\mathrm{lp}I}] |i\rangle, \label{eq:lpQRPAvac} \\
[10pt]
&& \mathcal{N}_{\mathrm{lp}I}^{\,2} = 
\prod_{K\pi} \langle i | \exp [v^{K\pi\dagger}_{\mathrm{lp}I}]
\exp [v^{K\pi}_{\mathrm{lp}I}] | i\rangle, \label {eq:lpQRPAvac_norm} \\
[10pt]
&& v^{K\pi}_{\mathrm{lp}I} = 
\sum_{\mu\nu\mu^\prime\nu^\prime} C^{\textrm{\scriptsize{lp}}I,K\pi}_{\mu\nu,-\mu^\prime -\nu^\prime} a^{i\dagger}_\mu a^{i\dagger}_\nu a^{i\dagger}_{-\mu^\prime} a^{i\dagger}_{-\nu^\prime}, \label{eq:lpQRPAvac_v} 
\end{eqnarray}
\vspace*{-10pt}
\begin{eqnarray}
&&(\mu \textrm{ and } \nu:\textrm{like-particles}, \mu<\nu; \nonumber \\
&&\, \mu^\prime \textrm{ and } {\nu^\prime}:\textrm{like-particles}, \mu^\prime<\nu^\prime), \nonumber
\end{eqnarray}
\begin{eqnarray}
&&O^\dagger_{B^{K\pi}_{\mathrm{lp}I}} = 
\sum_{\mu\nu} ( X^{\textrm{\scriptsize{lp}}I,K\pi}_{B,\mu\nu} 
a^{i\dagger}_\mu a^{i\dagger}_\nu - Y^{\textrm{\scriptsize{lp}}I,K\pi}_{B,{-\mu}{-\nu}} a^i_{-\nu} a^i_{-\mu}), \label{eq:Olpdagger} \\
&&(\mu \textrm{ and } \nu:\textrm{like-particles}, \mu<\nu), \nonumber
\end{eqnarray}
\begin{eqnarray}
C^{\textrm{\scriptsize{lp}}I,K\pi}_{\mu\nu,{-\mu^\prime}{-\nu^\prime}} &=& 
\frac{1}{1+\delta_{K0}}\sum_B Y^{\textrm{\scriptsize{lp}}I,K\pi\ast}_{B,{-\mu^\prime} {-\nu^\prime}} \nonumber \\
&& \times \left(\frac{1}{ X^{\textrm{\scriptsize{lp}}I,K\pi\ast} }\right)_{B,\mu\nu}. \label{eq:correlation_coeff_lp}
\end{eqnarray}
It is assumed that the quasiparticle states are ordered for using the notation of $\mu<\nu$. 
The condition of the good quantum numbers for the two-quasiparticle pairs, Eqs.~(\ref{eq:K_2qp}) and (\ref{eq:pi_2qp}), are also applied to Eq.~(\ref{eq:lpQRPAvac_v}). The equations analogous to Eqs.~(\ref{eq:lpQRPAvac})$-$(\ref{eq:lpQRPAvac_v}) can be written on the basis of the final state.

\subsection{\label{sec:extension} Extension and Test Term}
Now, the QRPA ground state with both the pnQRPA and lpQRPA correlations is defined as 
\begin{eqnarray}
|I\rangle = \frac{1}{\mathcal{N}_{\mathrm{pn}I}\mathcal{N}_{\mathrm{lp}I}} \prod_{K\pi} \exp [v^{K\pi}_{\mathrm{pn}I}] \exp [v^{K\pi}_{\mathrm{lp}I}]|i\rangle . \label{eq:QRPAvac}
\end{eqnarray}
I derive a test term for investigating the approximate commutability of the operators with the different origins. 
By using the above equations, it is possible to derive an expansion-truncation approximation
\begin{eqnarray}
\langle B^{K\pi}_{\mathrm{pn}I} |c^\dagger_p c_n |I\rangle
&\simeq& 
\langle I_{\textrm{\scriptsize{pn}}}|O_{B^{K\pi}_{\mathrm{pn}I}} c^\dagger_p c_n |I_{\textrm{\scriptsize{pn}}}\rangle \nonumber \\
&&+\frac{1}{ \mathcal{N}_{\mathrm{pn}I}^{\,2}}
\langle i| O_{B^{K\pi}_{\mathrm{pn}I}} c^\dagger_p c_n v^{K\pi}_{\mathrm{lp}I} |i\rangle .
\label{eq:transition_density_expanded}
\end{eqnarray}
The first term of the right-hand side is the zeroth-order term with respect to $v^{K\pi}_{\mathrm{lp}I}$. 
The second term is the cross term of the operators associated with  the pnQRPA and lpQRPA and linear with respect to 
$v^{K \pi}_{\mathrm{lp}I}$. 
If $v^{K \pi}_{\mathrm{lp}I}$ commutes with $O_{B^{K\pi}_{\mathrm{pn}I}} c^\dagger_p c_n$, that term vanishes. 
The factor $1/\mathcal{N}_{\mathrm{lp}I}^{\,2}$ does not have the linear term. 
The term 
$\langle i| v^{K\pi\dagger}_{\mathrm{lp}I} O_{B^{K\pi}_{\mathrm{pn}I}} c^\dagger_p c_n  |i\rangle$ 
is ignored because this is the second-order term with respect to the backward amplitudes. 
The terms $\langle i| O_{B^{K\pi}_{\mathrm{pn}I}} c^\dagger_p c_n v^{K^\prime\pi^\prime}_{\mathrm{lp}I} |i\rangle$ 
with $(K^\prime\pi^\prime)\neq (K\pi)$ are not included because of the commutability of the QRPA-order operators with different $(K\pi)$. 
The equations based on $|F\rangle$ are obtained analogously. 

It is relevant to this paper whether the commutability of the operators of the different QRPA in  $M^{(\mathrm{dcc})}$ is a good approximation. 
Thus, the contribution of the cross term is tested by calculating  
\begin{eqnarray}
\delta M^{(\textrm{\scriptsize{dcc}})} = 
M^{(\textrm{\scriptsize{dcc}})}_1 - M^{(\textrm{\scriptsize{dcc}})}_0, 
\label{eq:dMdcc_dcc_path}
\end{eqnarray}
\vspace{5pt}
\begin{eqnarray}
M_1^{(\textrm{\scriptsize{dcc}})} &=& \sum_{pp^\prime nn^\prime} \langle pp^\prime | V(\bm{r}) | nn^\prime \rangle \nonumber \\
&&\times\sum_{ B^{K\pi}_{\mathrm{pn}F} }
\sum_{ B^{K\pi}_{\mathrm{pn}I} }
\bigg\{
\langle F_{\textrm{\scriptsize{pn}}} | c^\dagger_{p^\prime} c_{n^\prime} 
O^\dagger_{B^{K\pi}_{\mathrm{pn}F}}
| F_{\textrm{\scriptsize{pn}}} \rangle \nonumber \\
&&+ \frac{1}{ \mathcal{N}^{\,2}_{\mathrm{pn}F} }
\langle f | v^{ K\pi\dagger }_{ \mathrm{lp}F }
c^\dagger_{p^\prime} c_{n^\prime} 
O^\dagger_{B^{K\pi}_{\mathrm{pn}F}} | f \rangle 
\bigg\} \nonumber \\
&&\times\langle B^{K\pi}_{\mathrm{pn}F} | B^{K\pi}_{\mathrm{pn}I} \rangle 
\bigg\{ \langle I_{\textrm{\scriptsize{pn}}}|O_{B^{K\pi}_{\mathrm{pn}I}} c^\dagger_p c_n |I_{\textrm{\scriptsize{pn}}}\rangle \nonumber \\
&&+\frac{1}{ \mathcal{N}_{\mathrm{pn}I}^{\,2} }
\langle i| O_{B^{K\pi}_{\mathrm{pn}I}} c^\dagger_p c_n v^{K\pi}_{\mathrm{lp}I} |i\rangle 
\bigg\}, 
\label{eq:Mdcc1_dcc_path}
\end{eqnarray}
\vspace{5pt}
\begin{eqnarray}
M_0^{(\textrm{\scriptsize{dcc}})} &=& \sum_{pp^\prime nn^\prime} \langle pp^\prime | V(\bm{r}) | nn^\prime \rangle \nonumber \\
&&\times\sum_{ B^{K\pi}_{\mathrm{pn}F} }
\sum_{ B^{K\pi}_{\mathrm{pn}I} }
\langle F_{\textrm{\scriptsize{pn}}} | c^\dagger_{p^\prime} c_{n^\prime} 
O^\dagger_{B^{K\pi}_{\mathrm{pn}F}}
| F_{\textrm{\scriptsize{pn}}} \rangle \nonumber \\
&&\times\langle B^{K\pi}_{\mathrm{pn}F} | B^{K\pi}_{\mathrm{pn}I} \rangle 
\langle I_{\textrm{\scriptsize{pn}}}|O_{B^{K\pi}_{\mathrm{pn}I}} c^\dagger_p c_n |I_{\textrm{\scriptsize{pn}}}\rangle . 
\label{eq:Mdcc0_dcc_path}
\end{eqnarray}
The test term $\delta M^\mathrm{(dcc)}$ includes the lowest-order contribution of the cross term to the transition density; only this one is tested because the cross-term calculation is rather costly computationally. 
For the explicit equations of 
$\langle f | v^{ K\pi\dagger }_{ \mathrm{lp}F } c^\dagger_{p^\prime} c_{n^\prime} 
O^\dagger_{B^{K\pi}_{\mathrm{pn}F}} | f \rangle $
and
$\langle i| O_{B^{K\pi}_{\mathrm{pn}I}} c^\dagger_p c_n v^{K\pi}_{\mathrm{lp}I} |i\rangle $, see Ref.~\cite{Ter16}. 
\\
\\

\begin{widetext}
\subsection{\label{sec:overlap}Overlap}
The approximate calculation of the overlap $\langle B^{K\pi}_{\mathrm{pn}F} |  B^{K\pi}_{\mathrm{pn}I} \rangle$ was developed and tested in Refs.~\cite{Ter12,Ter13}. According to that  study, I use the approximation 
\begin{eqnarray}
\lefteqn{ \langle B^{K\pi}_{\mathrm{pn}F} |  B^{K\pi}_{\mathrm{pn}I} \rangle } \nonumber \\
&\simeq&
\frac{1}{ \mathcal{N}_{ \mathrm{lp}I } \mathcal{N}_{ \mathrm{lp}F } }
\frac{1}{ \mathcal{N}_{\mathrm{pn}I} \mathcal{N}_{\mathrm{pn}F} }
\bigg\{ \langle f | O_{B^{ K\pi }_{\mathrm{pn}F} } O^\dagger_{B^{ K\pi }_{\mathrm{pn}I} } |i\rangle + \Big( \langle f | v^{K\pi\dagger}_{\mathrm{pn}F} O_{B^{K\pi}_{\mathrm{pn}F} } O^\dagger_{B^{ K\pi }_{\mathrm{pn}I} } |i\rangle 
+ \langle f | O_{B^{ K\pi }_{\mathrm{pn}F} } O^\dagger_{B^{ K\pi }_{\mathrm{pn}I} } v^{K\pi}_{\mathrm{pn}I} |i\rangle \Big) \bigg\}, \label{eq:overlap_approx}
\end{eqnarray}
\begin{eqnarray}
\langle f | O_{B^{ K\pi }_{\mathrm{pn}F} } O^\dagger_{B^{ K\pi }_{\mathrm{pn}I} } |i\rangle
 &=& \sum_{\mu<\nu} X^{\textrm{\scriptsize{pn}}F,K\pi \ast}_{B,\mu\nu} 
\sum_{\mu^\prime<\nu^\prime} X^{\textrm{\scriptsize{pn}}I,K\pi }_{B,\mu^\prime\nu^\prime} 
 \langle f | a^f_\nu a^f_\mu a^{i\dagger}_{\mu^\prime} a^{i\dagger}_{\nu^\prime}|i\rangle,
\end{eqnarray}
\begin{eqnarray}
\lefteqn{ \langle f | v^{K\pi\dagger}_{\mathrm{pn}F} O_{B^{ K\pi }_{\mathrm{pn}F} } O^\dagger_{B^{ K\pi }_{\mathrm{pn}I} } |i\rangle } \nonumber \\
&=& \sum_{\mu\nu\mu^\prime\nu^\prime} \sum_{\mu_1<\nu_1} \sum_{\mu_2<\nu_2}
C^{\textrm{\scriptsize{pn}}F,K\pi\ast}_{\mu\nu,\mu^\prime\nu^\prime}
X^{\textrm{\scriptsize{pn}}F,K\pi \ast}_{B,\mu_1\nu_1}
X^{\textrm{\scriptsize{pn}}I,K\pi }_{B,\mu_2\nu_2} 
\langle f | a^f_{\nu^\prime} a^f_{\mu^\prime} a^f_{\nu} a^f_{\mu} a^f_{\nu_1} a^f_{\mu_1}
a^{i\dagger}_{\mu_2} a^{i\dagger}_{\nu_2}|i\rangle 
-\sum_{\mu\nu} \sum_{\mu_1<\nu_1} \sum_{\mu_2<\nu_2}
Y^{\textrm{\scriptsize{pn}}F,K\pi \ast}_{B,{-\mu_1}{-\nu_1}}
X^{\textrm{\scriptsize{pn}}I,K\pi }_{B,\mu_2\nu_2} \nonumber \\
&&\times\Big\{ C^{\textrm{\scriptsize{pn}}F,K\pi\ast}_{-\nu_1-\mu_1,\mu\nu} 
-C^{\textrm{\scriptsize{pn}}F,K\pi\ast}_{-\mu_1-\nu_1,\mu\nu} 
+C^{\textrm{\scriptsize{pn}}F,K\pi\ast}_{\mu\nu,-\nu_1-\mu_1}
-C^{\textrm{\scriptsize{pn}}F,K\pi\ast}_{\mu\nu,-\mu_1-\nu_1} 
+C^{\textrm{\scriptsize{pn}}F,K\pi\ast}_{-\nu_1\nu,-\mu_1\mu} 
-C^{\textrm{\scriptsize{pn}}F,K\pi\ast}_{-\mu_1\nu,-\nu_1\mu} 
-C^{\textrm{\scriptsize{pn}}F,K\pi\ast}_{-\nu_1\nu,\mu-\mu_1} 
+C^{\textrm{\scriptsize{pn}}F,K\pi\ast}_{-\mu_1\nu,\mu-\nu_1} \nonumber \\
&&+C^{\textrm{\scriptsize{pn}}F,K\pi\ast}_{\mu-\nu_1,-\mu_1\nu} 
-C^{\textrm{\scriptsize{pn}}F,K\pi\ast}_{\mu-\mu_1,-\nu_1\nu}
-C^{\textrm{\scriptsize{pn}}F,K\pi\ast}_{\mu-\nu_1,\nu-\mu_1}
+C^{\textrm{\scriptsize{pn}}F,K\pi\ast}_{\mu-\mu_1,\nu-\nu_1} \Big\}
\langle f | a^f_\mu a^f_\nu a^{i\dagger}_{\mu_2} a^{i\dagger}_{\mu_2} | i \rangle.
\end{eqnarray}
\begin{eqnarray}
\lefteqn{ \langle f | O_{B^{ K\pi }_{\mathrm{pn}F} } O^\dagger_{B^{ K\pi }_{\mathrm{pn}I} } v^{K\pi}_{\mathrm{pn}I} |i\rangle } \nonumber \\
&=& \sum_{\mu<\nu} \sum_{\mu^\prime<\nu^\prime} \sum_{\mu_1\nu_1\mu_2\nu_2} 
X^{\textrm{\scriptsize{pn}}F,K\pi \ast}_{B,\mu\nu}
X^{\textrm{\scriptsize{pn}}I,K\pi }_{B,\mu^\prime\nu^\prime} 
C^{\textrm{\scriptsize{pn}}I,K\pi}_{\mu_1\nu_1,\mu_2\nu_2}
\langle f | a^f_{\nu} a^f_{\mu} a^{i\dagger}_{\mu^\prime} a^{i\dagger}_{\nu^\prime} a^{i\dagger}_{\mu_1} a^{i\dagger}_{\nu_1}
a^{i\dagger}_{\mu_2} a^{i\dagger}_{\nu_2}|i\rangle 
-\sum_{\mu<\nu} \sum_{\mu^\prime<\nu^\prime} \sum_{\mu_1\mu_2} 
X^{\textrm{\scriptsize{pn}}F,K\pi \ast}_{B,\mu\nu} 
Y^{\textrm{\scriptsize{pn}}I,K\pi }_{B,{-\mu^\prime}{-\nu^\prime}}
\nonumber \\
&&\times\Big\{ 
-C^{\textrm{\scriptsize{pn}}I,K\pi}_{\mu_1\mu_2,-\nu^\prime-\mu^\prime} 
+C^{\textrm{\scriptsize{pn}}I,K\pi}_{\mu_1\mu_2,-\mu^\prime-\nu^\prime} 
-C^{\textrm{\scriptsize{pn}}I,K\pi}_{-\nu^\prime-\mu^\prime,\mu_1\mu_2}
+C^{\textrm{\scriptsize{pn}}I,K\pi}_{-\mu^\prime-\nu^\prime,\mu_1\mu_2}
-C^{\textrm{\scriptsize{pn}}I,K\pi}_{\mu_1-\mu^\prime,\mu_2-\nu^\prime} 
+C^{\textrm{\scriptsize{pn}}I,K\pi}_{\mu_1-\nu^\prime,\mu_2-\mu^\prime} 
+C^{\textrm{\scriptsize{pn}}I,K\pi}_{\mu_1-\mu^\prime,-\nu^\prime\mu_2}  \nonumber \\
&&-C^{\textrm{\scriptsize{pn}}I,K\pi}_{\mu_1-\nu^\prime,-\mu^\prime\mu_2}
+C^{\textrm{\scriptsize{pn}}I,K\pi}_{-\mu^\prime\mu_1,\mu_2-\nu^\prime} 
-C^{\textrm{\scriptsize{pn}}I,K\pi}_{-\nu^\prime\mu_1,\mu_2-\mu^\prime}
-C^{\textrm{\scriptsize{pn}}I,K\pi}_{-\mu^\prime\mu_1,-\nu^\prime\mu_2}
+C^{\textrm{\scriptsize{pn}}I,K\pi}_{-\nu^\prime\mu_1,-\mu^\prime\mu_2} \Big\}
 \langle f | a^f_\nu a^f_\mu a^{i\dagger}_{\mu_1} a^{i\dagger}_{\mu_2} | i \rangle.
\end{eqnarray}
The normalization factor is also calculated by an expansion-truncation approximation. 
\begin{eqnarray}
\mathcal{N}_{\mathrm{pn}I} \simeq 
\bigg[ 1 + \sum_{K\pi} \Big\{ \langle i| v^{K\pi\dagger}_{\mathrm{pn}I}
v^{K\pi}_{\mathrm{pn}I}|i\rangle 
+\frac{1}{4} \langle i| (v^{K\pi\dagger}_{\mathrm{pn}I})^2
(v^{K\pi}_{\mathrm{pn}I})^2|i\rangle \Big\} \bigg]^{1/2}, 
\label{eq:norm_I}
\end{eqnarray}
\begin{eqnarray}
\langle i| v^{K\pi\dagger}_{\mathrm{pn}I}
\lefteqn{v^{K\pi}_{\mathrm{pn}I}|i\rangle } \nonumber \\
&=& (1+\delta_{K0}) \textrm{Tr} (C^{\textrm{\scriptsize{pn}}I,K\pi} C^{\textrm{\scriptsize{pn}}I,K\pi\dagger}) 
+\sum_{\mu\nu\mu^\prime\nu^\prime} C^{\textrm{\scriptsize{pn}}I,K\pi\ast}_{\mu\nu,\mu^\prime\nu^\prime} \Big( 
-C^{\textrm{\scriptsize{pn}}I,K\pi}_{\mu^\prime\mu,\nu^\prime\nu} 
+C^{\textrm{\scriptsize{pn}}I,K\pi}_{\mu^\prime\mu,\nu\nu^\prime}
-C^{\textrm{\scriptsize{pn}}I,K\pi}_{\nu^\prime\mu,\nu\mu^\prime} 
+C^{\textrm{\scriptsize{pn}}I,K\pi}_{\nu^\prime\mu,\mu^\prime\nu} \nonumber \\
&&
+C^{\textrm{\scriptsize{pn}}I,K\pi}_{\mu\mu^\prime,\nu^\prime\nu} 
-C^{\textrm{\scriptsize{pn}}I,K\pi}_{\mu\mu^\prime,\nu\nu^\prime} 
+C^{\textrm{\scriptsize{pn}}I,K\pi}_{\mu\nu^\prime,\nu\mu^\prime} 
-C^{\textrm{\scriptsize{pn}}I,K\pi}_{\mu\nu^\prime,\mu^\prime\nu}
+C^{\textrm{\scriptsize{pn}}I,K\pi}_{\mu^\prime\nu,\nu^\prime\mu} 
-C^{\textrm{\scriptsize{pn}}I,K\pi}_{\mu^\prime\nu,\mu\nu^\prime}
+C^{\textrm{\scriptsize{pn}}I,K\pi}_{\nu^\prime\nu,\mu\mu^\prime} 
-C^{\textrm{\scriptsize{pn}}I,K\pi}_{\nu^\prime\nu,\mu^\prime\mu} \nonumber \\
&&
-C^{\textrm{\scriptsize{pn}}I,K\pi}_{\nu\mu^\prime,\nu^\prime\mu} 
+C^{\textrm{\scriptsize{pn}}I,K\pi}_{\nu\mu^\prime,\mu\nu^\prime}
-C^{\textrm{\scriptsize{pn}}I,K\pi}_{\nu\nu^\prime,\mu\mu^\prime} 
+C^{\textrm{\scriptsize{pn}}I,K\pi}_{\nu\nu^\prime,\mu^\prime\mu} \Big),
\end{eqnarray}
where $C^{\textrm{\scriptsize{pn}}I,K\pi}$ is the matrix of which the matrix elements are 
$C^{\textrm{\scriptsize{pn}}I,K\pi}_{\mu\mu^\prime,\nu\nu^\prime}$, and the two-quasiparticle pairs are the row and column indexes. The higher-order term $(1/4)\langle i| (v^{K\pi\dagger}_{\mathrm{pn}I})^2
(v^{K\pi}_{\mathrm{pn}I})^2|i\rangle$ is approximated slightly simply;
\begin{eqnarray}
\frac{1}{4}
\langle i| (v^{K\pi\dagger}_{\mathrm{pn}I})^2
(v^{K\pi}_{\mathrm{pn}I})^2|i\rangle 
&\simeq&
\frac{1}{4} \Big\{ (2+6\delta_{K0})\big[\textrm{Tr} (C^{\textrm{\scriptsize{pn}}I,K\pi} C^{\textrm{\scriptsize{pn}}I,K\pi\dagger})\big]^2 
 + (2+14\delta_{K0})\textrm{Tr} (C^{\textrm{\scriptsize{pn}}I,K\pi} C^{\textrm{\scriptsize{pn}}I,K\pi\dagger})^2 \Big\}. 
\end{eqnarray}
\end{widetext}
The equations of $\mathcal{N}_{\mathrm{lp}I}$, $\mathcal{N}_{\mathrm{pn}F}$, and $\mathcal{N}_{\mathrm{lp}F}$ are derived in the analogous manner. 
It is necessary to calculate the generalized expectation value of product of the quasiparticle creation and annihilation operators as 
\begin{eqnarray}
\langle f | a^f_{\nu^\prime} a^f_{\mu^\prime} a^f_{\nu} a^f_{\mu} a^f_{\nu_1} a^f_{\mu_1}
a^{i\dagger}_{\mu_2} a^{i\dagger}_{\nu_2}|i\rangle. \label{eq:example_generalized_exp_value}
\end{eqnarray}
This term is calculated numerically using the generalized application of Wick's theorem. The technically important point is how to express the information of the contractions in the way applicable to the computation; for detail, see Appendix. 

\section{ \label{sec:application} Application }
\subsection{Technical procedure and parameters}
At the beginning, the HFB calculation is performed using the method of Refs.~\cite{Ter03,Bla05,Obe07}. The quasiparticle wave functions are represented by the B-spline on a non-uniform mesh of the cylindrical coordinate with the vanishing boundary condition. The top of the cylindrical box along the $z$ direction is at 20 fm from the origin, and the farthest  point from the origin to the direction perpendicular to the $z$ axis is also at 20 fm. Forty-two mesh points are used for each of the two intervals of the origin and the farthest points. I use the Skyrme (parameter set SkM$^\ast$ \cite{Bar82}) and volume contact pairing [proportional to $\delta(\boldsymbol{r})$ with no density dependence] interactions. 

I have so far applied the above method to $^{48}$Ca(initial)-$^{48}$Ti(final), $^{110}$Pd-$^{110}$Cd, $^{130}$Te-$^{130}$Xe, $^{136}$Xe-$^{136}$Ba, and $^{150}$Nd-$^{150}$Sm. 
Table \ref{tab:HFB_solution} shows the quadrupole deformation $\beta$ associated with the quadrupole moment and the lp pairing gaps of the HFB ground states. 
Table \ref{tab:strength_lp_pairing} shows the strengths of the lp pairing interactions determined by the usual method mentioned above and the pairing cutoff energy. 
If no cutoff energy is introduced, the effects of the pairing interaction diverge, thus the cutoff is essential. Since the pairing correlations are characterized by the localization of particles, the pairing  interactions are not expected to play an important role in a very high energy region with no bound or resonant particles. The quasiparticle levels with the possibility of the spacial localization are up to the absolute value of the single-particle potential depth. Thus, the smaller cutoff energies are used for the lighter nuclei in my calculation. 

For $^{48}$Ca, the strengths of the lp pairing interactions for $^{48}$Ti were used under the assumption that the adequate strength does not change significantly as $Z$ or $N$ changes slightly \cite{Ter18}. The usual method to determine the pairing-interaction strength is based on the assumption that the systematic odd-even mass difference  occurs solely because of the pairing correlations, therefore, this method is not appropriate for $Z$ or $N$ of the magic number. My HFB calculation yields a finite proton pairing gap for $^{48}$Ca (see Table \ref{tab:HFB_solution}). That HFB ground state is used in this study because the self-consistent calculation is the best possible way in the current circumstance; see Ref.~\cite{Ter18} for more detail. 

\begin{table}
\caption{\label{tab:HFB_solution} Properties of the HFB ground states of $^{48}$Ca, $^{48}$Ti, $^{110}$Pd, $^{110}$Cd, $^{130}$Te, $^{130}$Xe, $^{136}$Xe, $^{136}$Ba, $^{150}$Nd, and $^{150}$Sm. 
$\Delta_\mathrm{p}$ and $\Delta_\mathrm{n}$ are the average pairing gaps of the protons and neutrons, respectively, and $\beta$ denotes the quadrupole deformation parameter associated with the quadrupole moment \cite{Boh75}. }
\begin{ruledtabular}
\begin{tabular}{cccc}
Nucleus & $\Delta_\mathrm{p}$ (MeV) & $\Delta_\mathrm{n}$ (MeV) & $\beta$ \\
\hline\\[-10pt]
$^{48}$Ca & 1.731& 0& 0\\
$^{48}$Ti & 2.271& 1.731& 0\\
$^{110}$Pd & 1.397& 1.479& 0.252\\
$^{110}$Cd & 1.446& 1.401& 0.160\\
$^{130}$Te & $1.442$ & $1.359$ & $0$ \\
$^{130}$Xe & $1.692$ & $1.439$ & $0.112$ \\
$^{136}$Xe & $1.341$ & $0$ & $0$ \\
$^{136}$Ba & $1.641$ & $1.158$ & $0$ \\
$^{150}$Nd & 1.497& 0.914& 0.280\\
$^{150}$Sm & 1.882& 1.088& 0.206 
\end{tabular}
\end{ruledtabular}
\end{table}
\begin{table}
\caption{\label{tab:strength_lp_pairing} Strength of the volume contact pairing interactions for the protons ($G_\mathrm{pp}$) and the neutrons ($G_\mathrm{nn}$). The fourth column shows the cutoff energy for the pairing interaction relative to the bottom of the continuum region. }
\begin{ruledtabular}
\begin{tabular}{cccc}
\multirow{2}{*}{Nucleus} & $G_\mathrm{pp}$ & $G_\mathrm{nn}$ & Cutoff energy \\
            & (MeV$\,$fm$^3$)  & (MeV$\,$fm$^3$)  & (MeV) \\
\hline\\[-10pt]
$^{48}$Ca & $-$258.4 & $-$224.5 & 30 \\
$^{48}$Ti & $-$258.4 & $-$224.5 & 30\\
$^{110}$Pd & $-$224.3 & $-$185.5 & 50\\
$^{110}$Cd & $-$237.9 & $-$174.5 & 50\\
$^{130}$Te & $-$219.8 & $-$179.9 & 60\\
$^{130}$Xe & $-$219.8 & $-$181.8 & 60\\
$^{136}$Xe & $-$194.3 & $-$179.9 & 60\\
$^{136}$Ba & $-$200.5 & $-$189.3 & 60\\
$^{150}$Nd & $-$218.5 & $-$176.3 & 60\\
$^{150}$Sm & $-$218.5 & $-$181.6  & 60
\end{tabular}
\end{ruledtabular}
\end{table}

After the HFB solutions are obtained, the canonical-basis \cite{Rin80} wave functions are obtained by diagonalizing the nuclear one-body density. The  canonical-quasiparticle basis obtained by the BCS-like transformation of the canonical basis  is used for constructing the QRPA Hamiltonian matrix in so-called the matrix formulation \cite{Rin80}. Those operators of the canonical-quasiparticle basis are used for the equations in the previous section. The Hamiltonian is common for the HFB and QRPA calculations. 
The dimension of the two-canonical-quasiparticle basis is truncated by the method \cite{Ter15} on the basis of the occupation probability (if paired) or the HF single-particle energy (if the pairing gap vanishes). 
The dimension for the lpQRPA calculation is 60$\,$000$-$70$\,$000 for $K=0, 1$,  which have the spurious solutions, and less than 40$\,$000 for other $K$s; the larger $K$, the smaller dimension is used. I have performed the QRPA calculation up to $K=8$ (7 for $A=48$), for which the dimension was around 10$\,$000. For the pnQRPA calculation, the dimension is always less than 40$\,$000 because there is no spurious state. 
The QRPA solutions, i.e., $X$ and $Y$ amplitudes and eigen energies, are obtained by diagonalizing the QRPA Hamiltonian matrix \cite{Ter10}. 

The normalization factors of the QRPA ground states are necessary for the overlap calculations. 
Actually, the normalization factors diverge, if all the QRPA solutions are used for taking into account the QRPA correlations; see Eqs.~(\ref{eq:correlation_coeff_pn}) and (\ref{eq:correlation_coeff_lp}). It is known that the QRPA ground-state energy also diverges by including all the QRPA solutions for the calculation of this energy \cite{Mog10}. I used only the limited number of QRPA solutions with the largest backward-amplitude norms  
\begin{eqnarray}
\sum_{\mu\nu}(Y^{\mathrm{pn}I,K\pi}_{B,{-\mu}{-\nu}})^2 \textrm{\ or\ }
\sum_{\mu\nu}(Y^{\mathrm{lp}I,K\pi }_{B,{-\mu}{-\nu}})^2,
\label{eq:backward_norm}
\end{eqnarray}
of all the pn and lp QRPA solutions 
so as to reproduce the experimental binding energy of the initial nucleus. The same truncation method is applied to the final nucleus. Theoretically,  the binding energy is the summation of the contributions of the pn and lp QRPA correlations, however, it turned out in this truncation process that the necessary QRPA solutions were those of the lpQRPA. The low-energy lpQRPA solutions have more correlations than the pnQRPA solutions. 
Thus, neither $v^{K\pi}_{ \mathrm{pn}I }$ nor $v^{K\pi}_{ \mathrm{pn}F }$ is used in the overlap calculation, while the truncated $v^{K\pi}_{ \mathrm{lp}I }$ and $v^{K\pi}_{ \mathrm{lp}F }$ are used in the overlap calculation of the lpQRPA states except for $^{48}$Ca-$^{48}$Ti. 
The HFB calculation with SkM$^\ast$ overestimates the binding energies of $^{48}$Ca and $^{48}$Ti by 3$-$4 MeV, and the contributions of the QRPA solutions are positive to the binding energy. Thus, the overlap calculations of $^{48}$Ca-$^{48}$Ti are performed without the QRPA correlations of the ground states. 
The contributions of $v^{K\pi\dagger}_{ \mathrm{lp}F }$ and $v^{K\pi}_{ \mathrm{lp}I }$ to the unnormalized overlap of the lpQRPA states, 
\begin{eqnarray}
\langle f | \exp[{v^{K\pi\dagger}_{\mathrm{lp}F}}] O_{B^{K\pi}_{\mathrm{lp}F}} O^\dagger_{B^{K\pi}_{\mathrm{lp}I}}  
\exp[{v^{K\pi}_{\mathrm{lp}I}}] | i \rangle, 
\end{eqnarray}
do not diverge, if many QRPA solutions are used for constructing $v^{K\pi\dagger}_{\mathrm{lp}F}$ and $v^{K\pi}_{\mathrm{lp}I}$, because the initial and final HFB states have different configurations \cite{Ter18}. In fact, the effects of $v^{K\pi\dagger}_{\mathrm{lp}F}$ and $v^{K\pi}_{\mathrm{lp}I}$ to the unnormalized overlap is perturbative. $\mathcal{N}_{\mathrm{lp}I}$ and $\mathcal{N}_{\mathrm{lp}F}$ are also necessary in the overlap calculation of the pnQRPA states, except for $^{48}$Ca-$^{48}$Ti. The QRPA correlations have an effect to reduce the overlap. See Ref.~\cite{Ter15} for more detail. 

\subsection{ Double-$\bm{\beta}$ nuclear matrix element }

The half-life to the $0\nu\beta\beta$ decay $T^{(0\nu)}_{1/2}$, expected to be measured if the neutrino is a Majorana particle, is related to the effective neutrino mass $\langle m_\nu\rangle$ as 
\begin{eqnarray}
T^{(0\nu)}_{1/2} = \frac{ R^{(0\nu)}_{1/2} }{ \langle m_\nu \rangle^2 }. 
\end{eqnarray}
This $\langle m_\nu\rangle$ is defined by a transformation from the three neutrino eigen masses; see e.g., \cite{Doi85}. 
$R^{(0\nu)}_{1/2}$ is the quantity necessary to obtain theoretically and calculated by   
\begin{eqnarray}
R^{(0\nu)}_{1/2} = \frac{ m_e^2 c^4 }{ G_{0\nu} g_A^4 |M^{(0\nu)}|^2 }, \label{eq:R0v1/2}
\end{eqnarray}
where $G_{0\nu}$ is the phase-space factor \cite{Kot12} of the $0\nu\beta\beta$ decay, and the electron mass is denoted by $m_e c^2$. The effective axial-vector-current coupling is denoted by $g_A$. 
The $0\nu\beta\beta$-decay nuclear matrix element $M^{(0\nu)}$ is calculated in this paper according to  
\begin{eqnarray}
M^{(0\nu)} &=& M^{(0\nu)}_\mathrm{GT} - \frac{ g^2_V }{ g^2_A } M^\mathrm{(0\nu)}_\mathrm{F} , \label{eq:M0v}
\end{eqnarray}
with the vector-current coupling $g_V = 1$ and 
\begin{eqnarray}
M^{(0\nu)}_\mathrm{GT} &=& \sum_{ B^{K\pi}_{ \mathrm{pn}I },B^{K\pi}_{\mathrm{pn}F} } \sum_{pnp^\prime n^\prime}
\langle pp^\prime | V_{ \mathrm{GT} }^{(0\nu)} ( \bm{r};\bar{E}_B ) | nn^\prime \rangle \nonumber \\
&&\times \langle F | c^\dagger_p c_n | B^{K\pi}_{ \mathrm{pn}F } \rangle 
\langle B^{K\pi}_{ \mathrm{pn}F } | B^{K\pi}_{ \mathrm{pn}I } \rangle
\nonumber \\
&&\times\langle B^{K\pi}_{\mathrm{pn}I} | c^\dagger_{p^\prime} c_{n^\prime} | I \rangle , \label{eq:M0vGT}
\end{eqnarray}
\begin{eqnarray}
M^{(0\nu)}_\mathrm{F} &=& \sum_{ B^{K\pi}_{ \mathrm{pn}I },B^{K\pi}_{ \mathrm{pn}F } } \sum_{pnp^\prime n^\prime}
\langle pp^\prime | V_{ \mathrm{F} }^{(0\nu)}( \bm{r};\bar{E}_B ) | nn^\prime \rangle \nonumber \\
&&\times \langle F | c^\dagger_p c_n | B^{K\pi}_{\mathrm{pn}F} \rangle 
\langle B^{K\pi}_{\mathrm{pn}F} | B^{K\pi}_{\mathrm{pn}I} \rangle \nonumber \\
&&\times\langle B^{K\pi}_{\mathrm{pn}I} | c^\dagger_{p^\prime} c_{n^\prime} | I \rangle . \label{eq:M0vF}
\end{eqnarray} 
$M^{(0\nu)}_\mathrm{GT}$ and $M^{(0\nu)}_\mathrm{F}$ are the GT and Fermi components, respectively. These components are different in terms of the two-body transition operators 
\begin{eqnarray}
\lefteqn{ V_{ \mathrm{GT} }^{(0\nu)} ( \bm{r};\bar{E}_B ) } \nonumber \\
&=& h_+(r_{12};\bar{E}_B) \bm{\sigma}(1)\cdot \bm{\sigma}(2) \tau^-(1) \tau^-(2), \label{eq:v-potential sigma tau} 
\end{eqnarray}
\begin{eqnarray}
V_{ \mathrm{F} }^{(0\nu)} ( \bm{r};\bar{E}_B ) = 
h_+(r_{12};\bar{E}_B)  \tau^-(1) \tau^-(2), \label{eq:v-potential tau}
\end{eqnarray}
where $h_+(r_{12};\bar{E}_B$) is the neutrino potential, e.g., \cite{Doi85}, which is a function of the inter-nucleon distance $r_{12}$. The spin operator is denoted by $\bm{\sigma}$, and the operator $\tau^-$ changes the neutron to the proton. When Eq.~(\ref{eq:v-potential sigma tau}) is inserted to Eq.~(\ref{eq:M0vGT}), the operator $\bm{\sigma}(1)\tau^-(1)$ acts on $n$, and $\bm{\sigma}(2)\tau^-(2)$ acts on $n^\prime$. $V^{(0\nu)}_\mathrm{F}(\bm{r};\bar{E}_B)$ is used in the analogous manner. The neutrino potential arises from the neutrino exchange and depends on the energy of the intermediate state. It is known  that the intermediate-state energy can be approximated by an average value $\bar{E}_B$ (closure approximation \cite{Hor10,Sim11}) for the $0\nu\beta\beta$ decay. The tensor term, e.g., \cite{Eng17}, is omitted in my calculation of $M^{(0\nu)}$ because the contribution of that term is known to be small.  

\subsection{The test}
The test calculation discussed in Sec.~\ref{sec:extension} was performed for the component of $\delta M^\textrm{(dcc)}$  (\ref{eq:dMdcc_dcc_path}) with only the intermediate states of $(K\pi)=(2+)$, for simplicity, in $^{150}$Nd-$^{150}$Sm. The double-charge-change operator of the $0\nu\beta\beta$ decay
\begin{eqnarray}
V_{ \mathrm{GT} }^{(0\nu)} ( \bm{r};\bar{E}_B ) - \frac{ g^2_V }{ g^2_A } V_{ \mathrm{F} }^{(0\nu)} ( \bm{r};\bar{E}_B ),
\end{eqnarray}  
see Eqs.~(\ref{eq:v-potential sigma tau}) and (\ref{eq:v-potential tau}), was used for $V(\bm{r})$ in Eqs.~(\ref{eq:Mdcc1_dcc_path}) and (\ref{eq:Mdcc0_dcc_path}). The low-energy lpQRPA solutions with $(K\pi)=(2+)$ have the largest backward norms. 
It was found that 
$\delta M^\mathrm{(dcc)}[(K\pi)=(2+)]$ was only $-0.05$ \% of $M_0^\mathrm{(dcc)}[(K\pi)=(2+)]$  \cite{Ter16}. 
Thus, the operators associated with the different kind of QRPA commute with a very good accuracy. This commutability leads to 
\begin{eqnarray}
&&\langle F | c^\dagger_p c_n | B^{K\pi}_{ \mathrm{pn}F } \rangle \simeq \langle F_\mathrm{pn} | c^\dagger_p c_n | B^{K\pi}_{\mathrm{pn}F^\prime} \rangle , \nonumber \\[5pt]
&&\langle B^{K\pi}_{ \mathrm{pn}I } | c^\dagger_{p^\prime} c_{n^\prime} | I \rangle \simeq 
\langle B^{K\pi}_{ \mathrm{pn}I^\prime } | c^\dagger_{p^\prime} c_{n^\prime} | I_\mathrm{pn} \rangle , \label{eq:transition_density_matrix_element_approximated}
\end{eqnarray}
where
\begin{eqnarray}
&&| B^{K\pi}_{ \mathrm{pn}I^\prime } \rangle = O^\dagger_{B^{K\pi}_{ \mathrm{pn}I } } | I_\mathrm{pn}\rangle, \nonumber \\[5pt]
&&| B^{K\pi}_{ \mathrm{pn}F^\prime } \rangle = O^\dagger_{B^{K\pi}_{ \mathrm{pn}F } } | I_\mathrm{pn}\rangle, 
\end{eqnarray}
because of the normalization (\ref{eq:lpQRPAvac_norm}). Namely, the transition-density matrixes are calculated as usual. In the application of the lpQRPA, the analogous equations can be used. 

For the overlap of the two pnQRPA intermediate states, the normalization factor of the QRPA ground state needs the lpQRPA normalization factor $\mathcal{N}_{\mathrm{lp}I}$ or  $\mathcal{N}_{\mathrm{lp}F}$, thus, I have  
\begin{eqnarray}
\langle B^{K\pi}_{ \mathrm{pn}F } | B^{K\pi}_{ \mathrm{pn}I } \rangle \simeq 
\frac{ \langle B^{K\pi}_{ \mathrm{pn}F^\prime } | B^{K\pi}_{ \mathrm{pn}I^\prime } \rangle }{ \mathcal{N}_{\mathrm{lp}I} \mathcal{N}_{\mathrm{lp}F} }. \label{eq:overlap_approximated}
\end{eqnarray}
The approximations (\ref{eq:transition_density_matrix_element_approximated}) and (\ref{eq:overlap_approximated}) are used for the equations of $M^{(0\nu)}_\mathrm{GT}$ (\ref{eq:M0vGT}) and $M^{(0\nu)}_\mathrm{F}$ (\ref{eq:M0vF}) in the computation. 
Another expression of $M^{(0\nu)}$ by the equivalent pair transfer can be derived in the similar manner. 

\section{\label{sec:properties_strengths} Strengths of isoscalar pairing interaction by my method}
\subsection{\label{sec:IS_pairing_bb_NME} Effect of isoscalar pairing interaction to GT $\bm{0\nu\beta\beta}$ nuclear matrix element}
The GT component $M^{(0\nu)}_\mathrm{GT}$ is sensitive to the isoscalar pairing interaction and much less sensitive to the isovector pairing interaction because of the approximate isospin symmetry. 
The Fermi component $M^{(0\nu)}_\mathrm{F}$ is sensitive to the isovector pairing interaction and much less so to the isoscalar pairing interaction. Thus, I use $M^{(0\nu)}_\mathrm{GT}$ for $M^{(\mathrm{dcc})}$ of Eqs.~(\ref{eq:Mdcc_dcc_path}) and (\ref{eq:Mdcc_tpt_path}) for determining the strength of the isoscalar pairing interaction. 
This application is possible because of the closure approximation to the $0\nu\beta\beta$ nuclear matrix element.  
I show in Table \ref{tab:strength_pn_pairing} the strengths of the isoscalar ($G^\mathrm{IS}_\mathrm{pn}$) and isovector ($G^\mathrm{IV}_\mathrm{pn}$) pn-pairing volume contact interactions  determined by the method discussed in Sec.~\ref{sec:new_idea}. 
The value of $G^\mathrm{IS}_\mathrm{pn}$ ranges from $-180$ MeV$\,$fm$^3$ to $-50.0$ MeV$\,$fm$^3$. The range width is larger than that of $G^\mathrm{IV}_\mathrm{pn}$, thus, the nucleus dependence is large. The ratio of $G^\mathrm{IS}_\mathrm{pn}/G^\mathrm{IV}_\mathrm{pn}$ ranges from 0.25 to 0.83. Most values of the ratios of other methods are in the range of 1.05$-$1.9 (see Sec.~\ref{sec:introduction}), and the majority is around 1.5. My values are smaller than those of other methods. 

\begin{figure}[t]
\centerline{
\includegraphics[width=8.0cm]{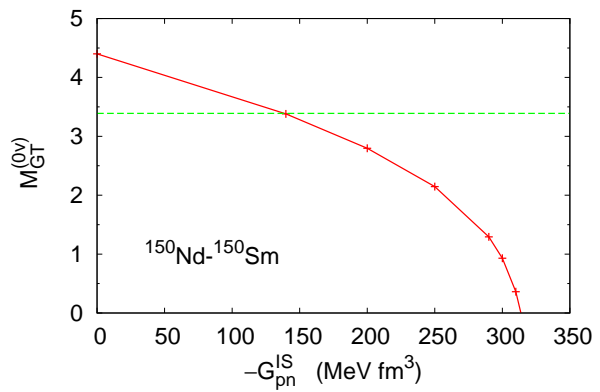} }
\caption{GT component $M^{(0\nu)}_\mathrm{GT}$ of 0$\nu\beta\beta$ nuclear matrix element for $^{150}$Nd-$^{150}$Sm calculated by the pnQRPA as a function of $G^\mathrm{IS}_\mathrm{pn}$ (solid line) and that calculated by the lpQRPA (dashed line). The latter is independent of the isoscalar pairing interaction, thus, it is drawn as a constant line.}
\label{fig:nme0vGT_150Nd-Sm}
\end{figure}

\begin{figure}[t]
\centerline{
\includegraphics[width=8.0cm]{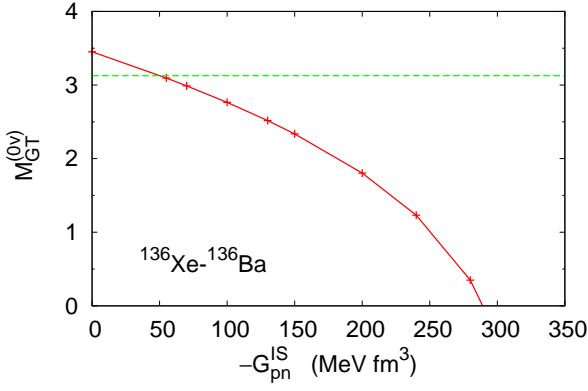} }
\caption{The same as Fig.~\ref{fig:nme0vGT_150Nd-Sm} but for $^{136}$Xe-$^{136}$Ba.}
\label{fig:nme0vGT_136Xe-Ba}
\end{figure}

\begin{figure}[t]
\centerline{
\includegraphics[width=8.0cm]{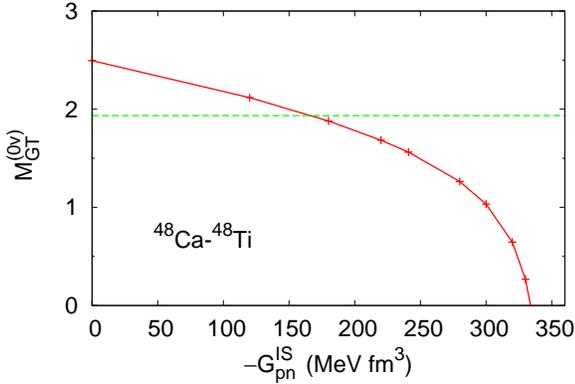} }
\caption{The same as Fig.~\ref{fig:nme0vGT_150Nd-Sm} but for $^{48}$Ca-$^{48}$Ti. }
\label{fig:nme0vGT_48Ca-Ti}
\end{figure}

I discuss how $G^\mathrm{IS}_\mathrm{pn}$ is found which satisfies the constraint. 
Three examples are shown in Figs.~\ref{fig:nme0vGT_150Nd-Sm}$-$\ref{fig:nme0vGT_48Ca-Ti} of the $G^\mathrm{IS}_\mathrm{pn}$ dependence of $M^{(0\nu)}_\textrm{GT}$ obtained by the pnQRPA calculations and the values by the lpQRPA calculations. 
The $G^\mathrm{IS}_\mathrm{pn}$ at the crossing point is that of my method. 
As Figs.~\ref{fig:nme0vGT_150Nd-Sm}$-$\ref{fig:nme0vGT_48Ca-Ti} show, the isoscalar pairing interaction has an effect to reduce the $\beta\beta$ nuclear matrix element $M^\mathrm{(0\nu)}_\mathrm{GT}$. In $^{150}$Nd-$^{150}$Sm,\footnote{In the calculation of $^{150}$Nd-$^{150}$Sm in  Ref.~\cite{Ter16}, I set $G^\mathrm{IV}_\mathrm{pn}=0$, and $M^{(0\nu)}$ was used for determining $G^\mathrm{IS}_\mathrm{pn}$. 
For unifying the systematic calculations, I recalculated this decay instance according to the method of this paper. The value of $M^{(0\nu)}$ of this paper corresponding to the previous one is $\simeq$4 \% larger.}
 $M^\mathrm{(0\nu)}_\mathrm{GT}$ diverges negatively at $G^\mathrm{IS}_\mathrm{pn}$ = $-$350.0 MeV$
\,$fm$^3$, and a QRPA solution has the zero energy accompanied by the diverging forward and backward amplitudes. This breaking point of the QRPA is caused by the isoscalar pn-pair condensate, thus, the divergence of $M^\mathrm{(0\nu)}_\mathrm{GT}$ occurs in any examples. The QRPA solutions are continuously connected to those at the breaking point; this continuity partially explains the behavior of $M^\mathrm{(0\nu)}_\mathrm{GT}$. The non-trivial question is why the contribution of the isoscalar pairing interaction is negative. 

\begin{table}
\caption{\label{tab:strength_pn_pairing} Strengths of the isoscalar ($G_\mathrm{pn}^\mathrm{IS}$) and isovector ($G_\mathrm{pn}^\mathrm{IV}$) pn pairing interactions determined by the methods explained in Sec.~\ref{sec:new_idea}. $G_\mathrm{pn}^\mathrm{IS}$ is determined for a pair of nuclei, thus, the same strength is used for the two nuclei. The fourth column shows the ratio of $G_\mathrm{pn}^\mathrm{IS}$ to $G_\mathrm{pn}^\mathrm{IV}$.}
\begin{ruledtabular}
\begin{tabular}{cccc}
Nucleus & $G_\mathrm{pn}^\mathrm{IS}$ (MeV$\,$fm$^3$) & $G_\mathrm{pn}^\mathrm{IV}$ (MeV$\,$fm$^3$) & $G_\mathrm{pn}^\mathrm{IS}/G_\mathrm{pn}^\mathrm{IV}$\\
\hline\\[-10pt]
$^{48}$Ca  &  $-$180.0&  $-$241.4 & 0.75\\
$^{48}$Ti   &  $-$180.0&  $-$241.4 & 0.75\\
$^{110}$Pd &  $-$170.0&  $-$204.9 & 0.83\\
$^{110}$Cd &  $-$170.0& $-$206.2 & 0.82\\
$^{130}$Te &  $\;\;-$50.0&  $-$199.9 & 0.25\\
$^{130}$Xe &  $\;\;-$50.0&  $-$200.8 & 0.25\\
$^{136}$Xe &  $\;\;-$55.0&  $-$187.1 & 0.29\\
$^{136}$Ba &  $\;\;-$55.0&  $-$187.1 & 0.29\\
$^{150}$Nd &  $-$139.8&  $-$197.4 & 0.71\\
$^{150}$Sm &  $-$139.8& $-$200.1 & 0.70 
\end{tabular}
\end{ruledtabular}
\end{table}
%

Let me consider a simplified single-charge-change transition matrix element
\begin{eqnarray}
\langle B | c^\dagger_{p\mu} c_{n\nu} + c^\dagger_{p\nu} c_{n\mu} | I \rangle,
\label{eq:simple_beta_matrix_element}
\end{eqnarray}
where $\langle B|$ is the abbreviation of $\langle B^\mathrm{pn}_I|$ in Sec.~\ref{sec:new_idea}, and $\mu$ and $\nu$ denote the quasiparticle indexes other than the proton or neutron for convenience in this discussion.  
The GT transition is considered, thus, $\mu \neq \nu$. The HFB states $|i\rangle$ and $|f\rangle$ are assumed to be pair condensate. When the isoscalar pairing interaction is enhanced,
the component
\begin{eqnarray}
\left( c^\dagger_{p\mu} c^\dagger_{n\nu} - c^\dagger_{n\mu} c^\dagger_{p\nu} \right) | I \rangle, 
\end{eqnarray}
of $| B \rangle$ is enhanced in the pnQRPA. The pn-pair creation is interpreted as an approximation to include the pn-pairing correlations of $|B\rangle$ with breaking of the conservation of the particle number. The component of the transition matrix element (\ref{eq:simple_beta_matrix_element})
\begin{eqnarray}
\lefteqn{ \langle I | (c_{n\nu} c_{p\mu} - c_{p\nu} c_{n\mu}) (c^\dagger_{p\mu} c_{n\nu} + c^\dagger_{p\nu} c_{n\mu}) | I \rangle } \nonumber \\
&=& 
\langle I | c_{n\nu} c_{p\mu} c^\dagger_{p\nu} c_{n\mu} - c_{p\nu} c_{n\mu} c^\dagger_{p\mu} c_{n\nu} | I \rangle, 
\end{eqnarray}
does not vanish because $|I\rangle$ does not conserve the particle number.
The corresponding component of the transition matrix element from $|B\rangle$ to the final state $|F\rangle$ is given by 
\begin{eqnarray}
\lefteqn{ \langle F | (c^\dagger_{p\mu} c_{n\nu} + c^\dagger_{p\nu} c_{n\mu}) (c^\dagger_{p\mu} c^\dagger_{n\nu} - c^\dagger_{n\mu} c^\dagger_{p\nu}) | F \rangle } \nonumber \\
&=& \langle F | -c^\dagger_{p\mu} c_{n\nu} c^\dagger_{n\mu} c^\dagger_{p\nu} + c^\dagger_{p\nu} c_{n\mu} c^\dagger_{p\mu} c^\dagger_{n\nu} | F \rangle. 
\end{eqnarray}
The double-charge-change matrix element in Sec.~\ref{sec:new_idea} is the summation with respect to $|B\rangle$ of the product of Eq.~(\ref{eq:simple_beta_matrix_element}) and the analogous matrix element between $\langle F|$ and $|B\rangle$. It is assumed for simplicity that the overlap matrix is a unit matrix. Here, a cancellation occurs as
\begin{eqnarray}
\sum_B \langle F | c^\dagger_{p\nu} c_{n\mu} | B \rangle \langle B | c^\dagger_{p\nu} c_{n\mu} | I \rangle = 0.
\end{eqnarray}
Therefore, the terms contributing to the summation are seen to be 
\begin{eqnarray}
&&-\langle F | c^\dagger_{p\mu} c_{n\nu}c^\dagger_{n\mu} c^\dagger_{p\nu} | F \rangle \langle I | c_{n\nu} c_{p\mu} c^\dagger_{p\nu} c_{n\mu} | I \rangle \nonumber \\
&&- \langle F | c^\dagger_{p\nu} c_{n\mu} c^\dagger_{p\mu} c^\dagger_{n\nu} | F \rangle
\langle I | c_{p\nu} c_{n\mu} c^\dagger_{p\mu} c_{n\nu} | I \rangle .
\label{eq:simple_isoscalar_dcc_matrix_element}
\end{eqnarray}
Analogously, the corresponding component including the intermediate states of the isovector excitation 
$c^\dagger_{p\mu} c^\dagger_{n\nu} + c^\dagger_{n\mu} c^\dagger_{p\nu}$
is seen to be the same as Eq.~(\ref{eq:simple_isoscalar_dcc_matrix_element}) but with the opposite sign. 

If the isovector-pairing contribution is positive to the double-charge-change transition matrix element, that sign difference explains the negative contribution of the isoscalar pairing interaction. 
I checked this assumption by using the Fermi nuclear matrix element $M^{(0\nu)}_\mathrm{F}$ of the $0\nu\beta\beta$ decay (\ref{eq:M0vF}) for $^{48}$Ca-$^{48}$Ti with $G^\mathrm{IV}_\mathrm{pn}=0$ and $-241.43$ MeV$\,$fm$^3$  ($G^\mathrm{IS}_\mathrm{pn}=-180.0$ MeV$\,$fm$^3$). The Fermi matrix element turned out to be $-$0.791 ($G^\mathrm{IV}_\mathrm{pn}=0$) and $-$0.320 ($G^\mathrm{IV}_\mathrm{pn}=-241.43$ MeV$\,$fm$^3$). Thus, the contribution of the isovector pairing interaction is positive to that matrix element. The corresponding variation of the GT nuclear matrix element $M^{(0\nu)}_\mathrm{GT}$ is only 0.2 \%. As seen from the calculation results presented above, $M^{(0\nu)}_\mathrm{GT}$ and $M^{(0\nu)}_\mathrm{F}$ have the opposite signs. Thus, an important reason for the characteristic effect of the isoscalar pairing interaction is the sign difference included in  the isoscalar and isovector operators. It is implied that this effect is qualitatively independent of nucleus. 

My computer programs are developed assuming no pn-pair condensate. If the condensation occurs, the HFB ground state depends on $G^\mathrm{IS}_\mathrm{pn}$ and/or $G^\mathrm{IV}_\mathrm{pn}$. Thus, $M^{(0\nu)}_\mathrm{GT}$ calculated by the lpQRPA is not a constant in the condensate region. The behaviors of $M^{(0\nu)}_\mathrm{GT}$'s by the two QRPA in that region are unknown. 

\subsection{\label{sec:systematics_strength_isoscalar_pairing} Systematics of strengths of isoscalar pairing interaction}
\begin{figure}[t]
\centerline{
\includegraphics[width=8.0cm]{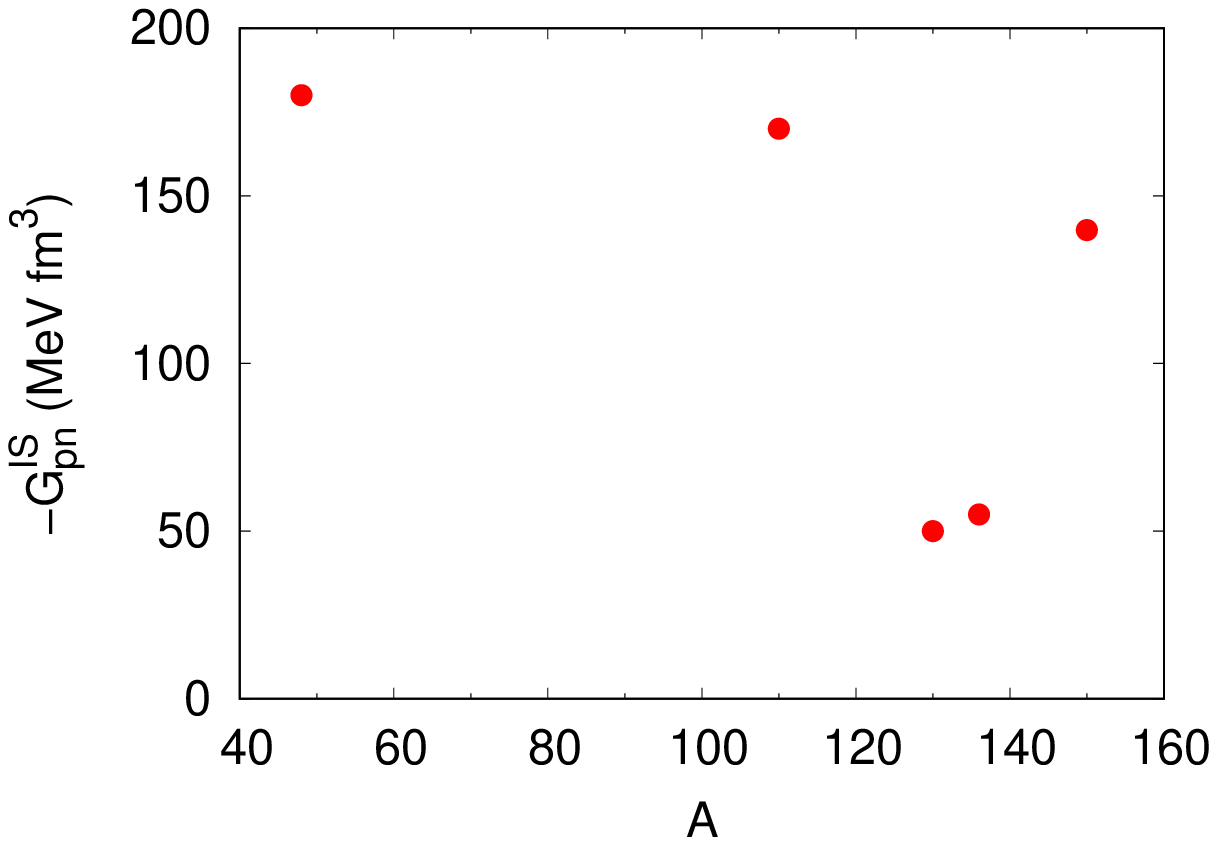} }
\caption{$-G^\mathrm{IS}_\mathrm{pn}$ as a function of $A$. The values are noted in Table \ref{tab:strength_pn_pairing}.}
\label{fig:a_gt0_paper}
\end{figure}

\begin{figure}[t]
\centerline{
\includegraphics[width=8.0cm]{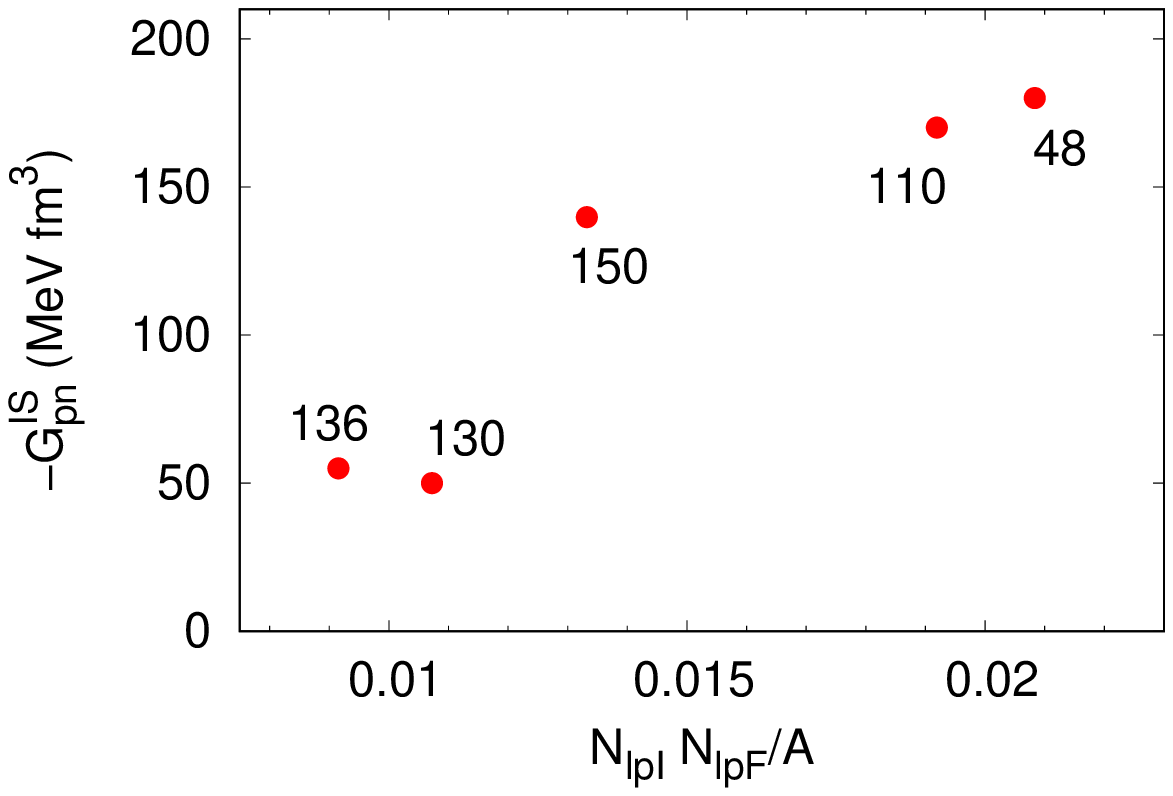} }
\caption{$-G^\mathrm{IS}_\mathrm{pn}$ as a function of $\mathcal{N}_{\mathrm{lp}I} \mathcal{N}_{\mathrm{lp}F}/A$ with $A$ inserted. For $\mathcal{N}_{\mathrm{lp}I}$ and $\mathcal{N}_{\mathrm{lp}F}$, see Sec.~\ref{sec:new_idea}.}
\label{fig:nrmoa_gt0_paper}
\end{figure}

\begin{figure}[t]
\centerline{
\includegraphics[width=8.0cm]{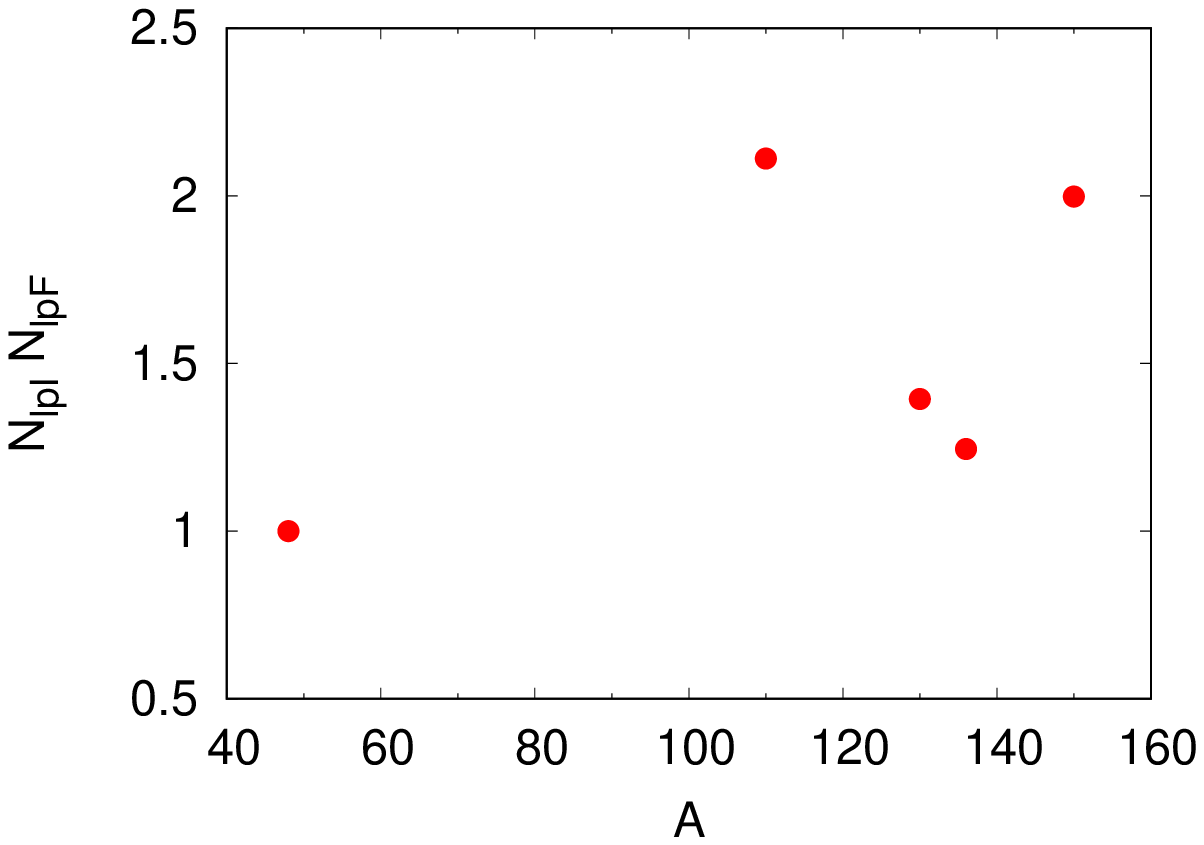} }
\caption{$\mathcal{N}_{\mathrm{lp}I} \mathcal{N}_{\mathrm{lp}F}$ as a function of $A$.}
\label{fig:a_nrm_paper}
\end{figure}

Figure \ref{fig:a_gt0_paper} is the illustration of the $A$ dependence of $G^\mathrm{IS}_\mathrm{pn}$ given by Table \ref{tab:strength_pn_pairing}. The figure shows two groups of values of $G^\mathrm{IS}_\mathrm{pn}$, however, a figure more useful for getting the physical insight is obtained by plotting $-G^\mathrm{IS}_\mathrm{pn}$ as a function of $\mathcal{N}_{\mathrm{lp}I} \mathcal{N}_{\mathrm{lp}F}/A$ (Fig.~\ref{fig:nrmoa_gt0_paper}); $G^\mathrm{IS}_\mathrm{pn}$ is approximately proportional to $\mathcal{N}_{\mathrm{lp}I} \mathcal{N}_{\mathrm{lp}F}/A$. For the normalization factors of the lpQRPA ground states $\mathcal{N}_{\mathrm{lp}I}$ and $\mathcal{N}_{\mathrm{lp}F}$, see Sec.~\ref{sec:new_idea}. The $A$ dependence of $\mathcal{N}_{\mathrm{lp}I} \mathcal{N}_{\mathrm{lp}F}$ is depicted in Fig.~\ref{fig:a_nrm_paper}. The comparison of Figs.~\ref{fig:a_gt0_paper} and \ref{fig:a_nrm_paper} seemingly indicates that $-G^\mathrm{IS}_\mathrm{pn}$ and $\mathcal{N}_{\mathrm{lp}I} \mathcal{N}_{\mathrm{lp}F}$ are correlated except for $^{48}$Ca. However, this nucleus fits the systematics in Fig.~\ref{fig:nrmoa_gt0_paper} well, thus, it is appropriate to consider $\mathcal{N}_{\mathrm{lp}I} \mathcal{N}_{\mathrm{lp}F}/A$.  

$1/A$ is the global $A$ dependence of the strength of the lp pairing interaction in the nuclear chart, known phenomenologically \cite{Boh75} and in a simple harmonic-oscillator estimation \cite{Bes69}. The $G^\mathrm{IS}_\mathrm{pn}$ includes this property. $\mathcal{N}_{\mathrm{lp}I}$ and $\mathcal{N}_{\mathrm{lp}F}$ reflect on the many-body correlations because these factors  deviate from 1 by the QRPA correlations. Qualitatively, the QRPA normalization factor is close to 1 for the magic nuclei and larger for midshell nuclei. The initial and final states of the $\beta\beta$ decay involve two proton and two neutron numbers. $^{48}$Ca-$^{48}$Ti has two magic numbers, and $^{136}$Xe-$^{136}$Ba has one. $^{130}$Te-$^{130}$Xe has the proton and neutron numbers close to the magic numbers, and $^{110}$Pd, $^{110}$Cd, $^{150}$Nd, and $^{150}$Sm are midshell nuclei. Thus, the $1/A$ dependence is modified in the manner of enhancing $G^\mathrm{IS}_\mathrm{pn}$ for the midshell nuclei. This midshell effect is more than perturbative, so that $G^\mathrm{IS}_\mathrm{pn}$ does not have a simple $A$ dependence. The pairing correlations are enhanced generally in the midshell nuclei compared to the closed-shell nuclei because the level density of the single particles around the Fermi surface is relevant. The $G^\mathrm{IS}_\mathrm{pn}$ reflects on this property, as the strength is determined by a mathematical equality reflecting on the lp pairing correlations; see Eq.~(\ref{eq:Mdcc_tpt_path}). 

\section{\label{sec:comparison_str_exp} Comparison of charge-change transition strength with experimental data}
\begin{figure*}[t]
\centerline{
\includegraphics[width=15.0cm]{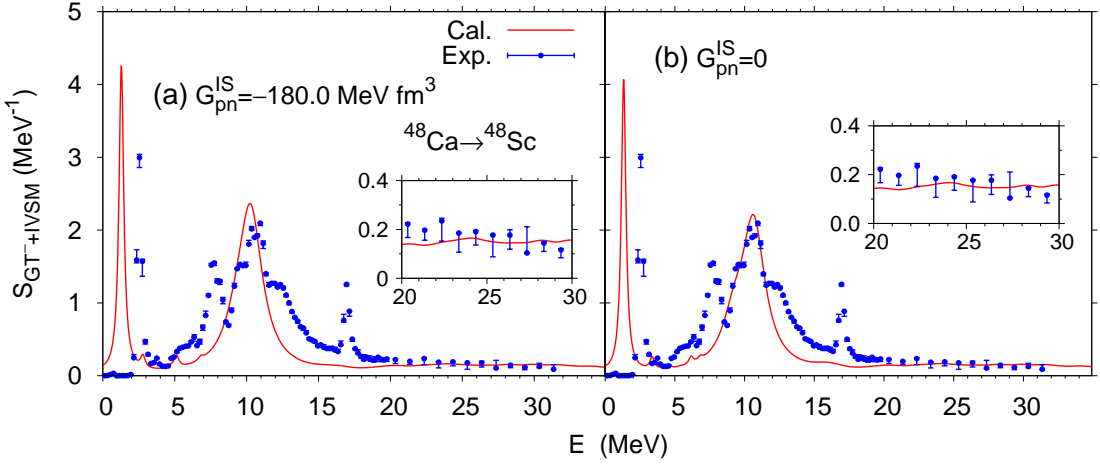} }
\caption{ Charge-change strength function by the GT and isovector spin monopole (IVSM) operator for transition from $^{48}$Ca to $^{48}$Ti. $E$ is the excitation energy of the final nucleus. The calculated result is shown by line, and the experimental data \cite{Yak09} are shown by isolated symbols; (a) the calculation was performed with $G^\mathrm{IS}_\mathrm{pn} = -180.0$ MeV$\,$fm$^3$, and (b) $G^\mathrm{IS}_\mathrm{pn} = 0$ was used. The inset is a magnification of a high-energy region. Figure (a) was taken from Ref.~\cite{Ter18}.}
\label{fig:strfn_mix_cmbd_48Ca}
\end{figure*}
The charge-change transition strengths have been obtained by experiments for some nuclei.  Figure \ref{fig:strfn_mix_cmbd_48Ca} shows the single-charge-change strength function of $J^\pi=1^+$ 
 for $^{48}$Ca$\rightarrow^{48}$Sc \cite{Yak09}, and corresponding my calculation is shown by lines. The  calculation result in the left panel was obtained with $G^\mathrm{IS}_\mathrm{pn}=-180.0$ MeV$\,$fm$^3$, and for that in the right panel $G^\mathrm{IS}_\mathrm{pn}=0$ was used. 
The transition operator used in the calculation is a linear combination of two operators 
\begin{eqnarray}
\bm{\sigma}\tau^- +\alpha r^2 \bm{\sigma}\tau^-, \label{eq:GT+isovector_spin_monopole_operator}
\end{eqnarray}
where the second term is the isovector spin monopole operator with a factor $\alpha$ so as to reproduce the experimental data in the tail region shown in the inset. For detail of this discussion, see Ref.~\cite{Ter18}. The only visible difference between the two calculations is the peak at $E=10$ MeV. However, the corresponding lowest-energy peaks are also slightly different; the energy of that peak is 1.289 MeV ($G^\mathrm{IS}_\mathrm{pn}=-180.0$ MeV$\,$fm$^3$) and 1.321 MeV ($G^\mathrm{IS}_\mathrm{pn}=0$), and the transition strength is 2.653 ($G^\mathrm{IS}_\mathrm{pn}=-180.0$ MeV$\,$fm$^3$) and 2.534 ($G^\mathrm{IS}_\mathrm{pn}=0$). The energy is decreased by 32 keV (2.4 \%) by that isoscalar pairing interaction, and the transition strength is increased by 4.7 \%. There are a couple of reasons for this smallness of the difference. Firstly, my value of $G^\mathrm{IS}_\mathrm{pn}$ is much smaller than the strength at the breaking point of the QRPA, and secondly, $^{48}$Ca is a doubly-magic nucleus. The problem in terms of the reproductivity of the experimental data is that the two major peaks are not reproduced simultaneously, and this discrepancy is not removed by the isoscalar pairing interaction. Probably this problem indicates the necessity of a better particle-hole interaction. It is also seen by comparing this result with Fig.~\ref{fig:nme0vGT_48Ca-Ti} that $M^\mathrm{(0\nu)}_\mathrm{GT}$ is more sensitive to the isoscalar pairing interaction than the single-charge-change transition strength. 

\begin{figure}[t]
\centerline{
\includegraphics[width=8.0cm]{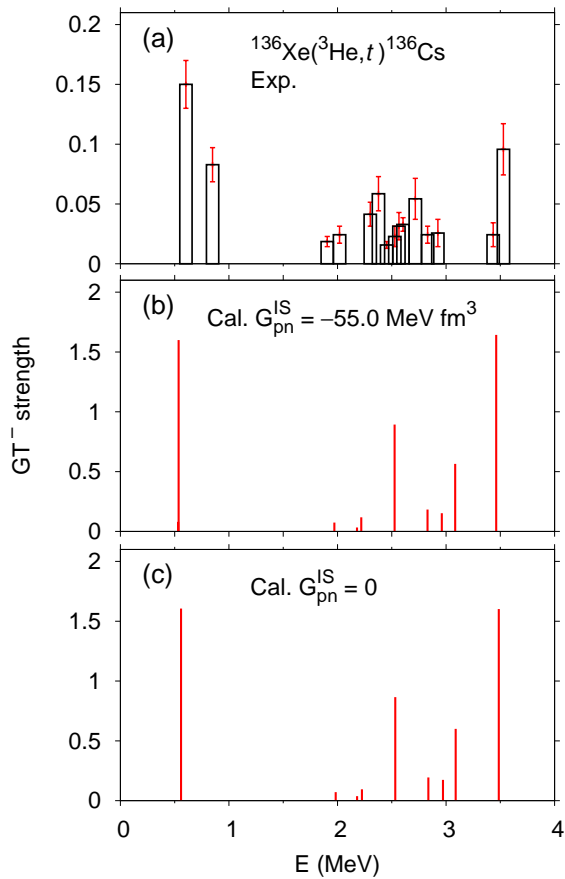} }
\caption{GT transition strength from $^{136}$Xe to $^{136}$Cs in a low-energy region; (a) experimental data deduced from cross section of ($^3$He,$t$) reaction \cite{Pup11}, (b) calculated GT strength with $G^\mathrm{IS}_\mathrm{pn}=-55.0$ MeV$\,$fm$^3$, and (c) the same as (b) but for $G^\mathrm{IS}_\mathrm{pn}=0$. Figure (b) was taken from Ref.~\cite{Ter19}.}
\label{fig:gtstr_cmbd_136Xe}
\end{figure}

Figure \ref{fig:gtstr_cmbd_136Xe} shows the GT transition strengths from $^{136}$Xe to $^{136}$Cs deduced from the charge-change reaction \cite{Pup11} and the calculated ones. The overall energy dependence of the strength distribution is reproduced by the calculation, however, the   calculated strengths are one order of magnitude smaller than the experimental data. In this $^{136}$Xe-$^{136}$Cs calculation, I did not include the isovector spin monopole operator because the data in the higher-energy region not shown in the figure are necessary for verifying the contribution of that operator.  It is an open question what can be learned from the data other than the energy dependence.

The calculated lowest-energy peak is at 5.161 MeV ($G^\mathrm{IS}_\mathrm{pn}=-55.0$ MeV$\,$fm$^3$) and 5.183 MeV ($G^\mathrm{IS}_\mathrm{pn}=0$), and the GT strength of that peak is 1.599 ($G^\mathrm{IS}_\mathrm{pn}=-55.0$ MeV$\,$fm$^3$) and 1.606 ($G^\mathrm{IS}_\mathrm{pn}=0$). The energy is decreased by 22 keV (0.4 \%) because of the isoscalar pairing interaction, and the GT strength is also decreased only by 0.4 \%. As anticipated from the relatively small strength of the isoscalar pairing interaction, its effect is small. The comparison of the data and the calculation seems to indicate  that this weak isoscalar pairing interaction is not a problem. 

\section{\label{sec:NME_0vbb} Calculation results of nuclear matrix elements of $\bm{0\nu\beta\beta}$ decay}
The appropriate value of the axial-vector-current coupling $g_A$ is a long-standing issue to the studies of the $\beta\beta$-decay nuclear matrix elements; see, e.g., Refs.~\cite{Eng17,Suh17b} for review and the discussion  below. The bare value of $g_A$ is 1.27, which is determined by the half-life of the decay of neutron. However, the theoretical calculations do not reproduce the measured half-lives of the nuclear $\beta$  decays with this value  systematically \cite{Bro85}. This is quite in contrast to the success that the systematics of many measured electric transition strengths can be reproduced  with the bare charge \cite{Ter08}. A method to determine the effective $g_A$ is to fit the measured half-lives of the $2\nu\beta\beta$ decays (see Sec.~\ref{sec:introduction}). This half-life is calculated by 
\begin{eqnarray}
T^{(2\nu)}_{1/2} = \frac{1}{ G^{(0)}_{2\nu} g_A^4 | M^{(2\nu)}|^2 },
\end{eqnarray}
where $G^{(0)}_{2\nu}$ is the phase-space factor of the $2\nu\beta\beta$ decay \cite{Kot12}. The nuclear matrix element of this decay $M^{(2\nu)}$ is defined 
\begin{eqnarray}
M^{(2\nu)} = \frac{ M^{(2\nu)}_\mathrm{GT} }{ \mu_0 } - \frac{g_V^2}{g_A^2} \frac{ M^{(2\nu)}_\mathrm{F} }{ \mu_{0F} }, \label{eq:M2v}
\end{eqnarray}
\begin{eqnarray}
\frac{ M^{(2\nu)}_\mathrm{GT} }{\mu_0} &=&\sum_{K=0,\pm 1} \sum_{ B_{\mathrm{pn}I}^{K+}, B_{\mathrm{pn}F}^{K+} } \frac{1}{\mu_B} \langle F | \tau^- (-)^K \sigma_{-K} | B_{\mathrm{pn}F}^{K+} \rangle \nonumber \\
&&\times \langle B_{\mathrm{pn}F}^{K+} | B_{\mathrm{pn}I}^{K+} \rangle
\langle B_{\mathrm{pn}I}^{K+} | \tau^- \sigma_K | I \rangle, \label{eq:M2v_GT}
\end{eqnarray}
\begin{eqnarray}
\frac{ M^{(2\nu)}_\mathrm{F} }{\mu_{0F}} &=& \sum_{ B_{ \mathrm{pn}I}^{0+}, B_{\mathrm{pn}F}^{0+} } \frac{1}{\mu_B} \langle F | \tau^- | B_{\mathrm{pn}F}^{0+} \rangle \langle B_{\mathrm{pn}F}^{0+} | B_{\mathrm{pn}I}^{0+} \rangle \nonumber \\
&& \times \langle B_{\mathrm{pn}I}^{0+} | \tau^- | I \rangle ,
\end{eqnarray}
\begin{eqnarray}
\mu_B = \frac{1}{m_e c^2} ( E_B - \bar{M} ).
\end{eqnarray}
$E_B$ is the abbreviation of the intermediate-state energy, that is, it is either that obtained from the initial state $E(B^{K\pi}_{\mathrm{pn}I})$ or that obtained from the final state $E(B^{K\pi}_{\mathrm{pn}F})$. 
$\bar{M}$ denotes the mean value of the masses of the initial and final nuclei. If the isospin symmetry is complete, the Fermi component vanishes. 
This is because the isospin of the final state is smaller than that of the initial state by 2 (in many nuclei, the ground state has $T$ = $|T_z|$), however, $\tau^-$ only changes $T_z$. 
The explicit intermediate-state-energy dependence of $M^{(2\nu)}$ cannot be neglected because this decay does not involve the virtual-neutrino propagator between the nucleons. Therefore, the closure approximation is not used in the $2\nu\beta\beta$-decay calculations. 
$M^{(2\nu)}$ is not used for $M^{\mathrm{(dcc)}}$ of the constraint to the effective interactions.
In the QRPA approach, the mean value of $E(B^{K\pi}_{\mathrm{pn}I})$ and $E(B^{K\pi}_{\mathrm{pn}F})$ have been used with the overlaps of all the possible combinations of $|B^{K\pi}_{\mathrm{pn}I}\rangle$ and $|B^{K\pi}_{\mathrm{pn}F}\rangle$, e.g., \cite{Sim09} and references therein. I have performed two calculations with $E_B=E(B^{K\pi}_{\mathrm{pn}I})$ and $E_B=E(B^{K\pi}_{\mathrm{pn}F})$. 
If the two results coincide, the QRPA approach is a good approximation; see below.  

In early days, $g_A=1.0$ has been often used because this value was the standard for the studies of the $\beta$ decays \cite{Bro85}. These days, the possibility of $g_A < 1.0$ is also investigated \cite{Fae08, Suh13, Juo05}. 
The reason is not established why the effective value for nuclei is smaller than the bare value. It is also a problem whether the effective $g_A$ for the $0\nu\beta\beta$ decay is different from that for the $2\nu\beta\beta$ decay. The former $g_A$ cannot be confirmed experimentally because there is no observed phenomenon caused by the neutrino-exchange interaction in a nucleus. 

The value of $g_V=1.0$ is used in most of the calculations for the $\beta\beta$ decays. My speculative reason is that the Fermi constant obtained from the muon decay and that from the isobaric analog Fermi decays of nuclei are very close to each other \cite{Boh69, Com83}, and $g_V=1$ is used in both methods. 
In the fundamental weak interaction at the quark level, e.g.,  \cite{Com83}, the values of $g_A$ and $g_V$ are both one. 

Table \ref{tab:0vbb_NME} shows $M^{(0\nu)}_\mathrm{GT}$ (\ref{eq:M0vGT}), $M^{(0\nu)}_\mathrm{F}$  (\ref{eq:M0vF}), $M^{(0\nu)}$ (\ref{eq:M0v}), and $R^{(0\nu)}_{1/2}$ (\ref{eq:R0v1/2}) for five decay instances. Multiple $M^{(0\nu)}$'s and $R^{(0\nu)}_{1/2}$'s are noted for every decay except for $^{110}$Pd-$^{110}$Cd. Under the uncertainty of $g_A$, it is a custom to present $M^{(0\nu)}$ with the $g_A$ reproducing the half-life of the $2\nu\beta\beta$ decay, denoted by $g_A(2\nu\beta\beta)$, and the bare value of $g_A$, e.g., \cite{Sim09}. The results with these two $g_A$ values seem to be referred to for speculating the likely range of the $M^{(0\nu)}$ and $R^{(0\nu)}_{1/2}$ because there is no theoretical guide line for using values smaller than $g_A(2\nu\beta\beta)$ or those larger than the bare value.  
Some decay instances have results with two $g_A(2\nu\beta\beta)$. This is because $M^{(2\nu)}$  depends on the set of $E(B^{K\pi}_{\mathrm{pn}I})$ or $E(B^{K\pi}_{\mathrm{pn}F})$. $^{48}$Ca-$^{48}$Ti and $^{136}$Xe-$^{136}$Ba have only one $g_A(2\nu\beta\beta)$ because these decay instances do not have that dependence on the set of the intermediate-state energies. Therefore, the QRPA approach to those decay instances is a good approximation. In Ref.~\cite{Ter19}, the validity was investigated in detail including the intermediate-state-energy set dependence for $^{136}$Xe-$^{136}$Ba and $^{130}$Te-$^{130}$Xe, and it turns out that the QRPA approach is more reliable in the former decay than the latter one. $^{110}$Pd-$^{110}$Cd does not have a result with $g_A(2\nu\beta\beta)$ because there is no experimental data of the $2\nu\beta\beta$ decay. 

If, for example, $\langle m_\nu\rangle$ is equal to 10 meV, the half-life of $^{136}$Xe to the $0\nu\beta\beta$ decay might be around 1.3$\times 10^{23}$ yr ($g_A=0.49)$ or 6.1$\times 10^{21}$ yr ($g_A=1.27)$. In either estimation, it is an extremely long half-life. The difference between the two estimations is a factor of 20 approximately, and those decays shown in Table \ref{tab:0vbb_NME} have the differences of a factor of $14$$-$$21$. This large difference is a problem for designing the experimental setup for the future. 
The average of the measured $T^{(2\nu)}_{1/2}$ is $(2.18\pm 0.05)\times 10^{21}$ yr \cite{Bar19}. 
The progress of the sensitivity of the experiments is remarkable. 

\begin{table}
\caption{\label{tab:0vbb_NME} $M^{(0\nu)}_\mathrm{GT}$, $M^{(0\nu)}_\mathrm{F}$, $M^{(0\nu)}$, $R^{(0\nu)}_{1/2}$, and $g_A$ of five $0\nu\beta\beta$-decay instances. The results are shown with the $g_A$ reproducing the experimental half-life of the $2\nu\beta\beta$ decay [$g_A(2\nu\beta\beta)$] and $g_A=1.27$  (bare value). Those $g_A$'s other than 1.27 are $g_A(2\nu\beta\beta)$. Two $g_A(2\nu\beta\beta)$'s are shown in two decay instances, in which $g_A(2\nu\beta\beta)$ depends on the set of intermediate-state energies; one is that obtained by the pnQRPA based on the initial state, and another is based on the final state. For the decay instances with only one $g_A(2\nu\beta\beta)$, there is no that dependence. $^{110}$Pd-$^{110}$Cd shows only the result with $g_A=1.27$ because there is no experimental data of the $2\nu\beta\beta$ decay. }
\begin{ruledtabular}
\begin{tabular}[t]{cccccc}
\multirow{3}{*}{ Decay } & \multirow{3}{*}{ $M^{(0\nu) }_\mathrm{GT}$ } & 
\multirow{3}{*}{ $M^{(0\nu)}_\mathrm{F}$ } & 
\multirow{3}{*}{ $M^{(0\nu) }$ } & 
$R^{(0\nu)}_{1/2}$ & \multirow{3}{*}{ $g_A$ } \\
 & & & & (10$^{13}$\,MeV$^2\,$ & \\[-3pt]
 & & & &  yr) & \\
\hline\\[-10pt]
$^{48}$Ca-$^{48}$Ti & 1.880 & $-$0.349 & 
$\left\{\begin{array}{c} 3.332 \\[-2pt]
2.096 
\end{array}\right.$ & 
$\left\{\begin{array}{c} 1.645 \\[-2pt]
0.092 
\end{array}\right.$ & 
$\left\{\begin{array}{c} 0.49 \\[-2pt]
1.27
\end{array}\right.$ \\[9pt]
$^{110}$Pd-$^{110}$Cd & 2.486 & $-$0.571 & 2.840 & 0.258 & 
1.27 \\[2pt]
$^{130}$Te-$^{130}$Xe & 
3.613 & $-$0.709 & 
$\left\{\begin{array}{c}6.692 \\[-2pt]
 5.400 \\[-2pt]
 4.053 
\end{array}\right.$ & 
$\left\{\begin{array}{c} 0.772 \\[-2pt]
0.400 \\[-2pt]
0.043  
\end{array}\right.$ & 
$\left\{\begin{array}{c} 0.48 \\[-2pt]
0.63 \\[-2pt]
1.27
\end{array}\right.$ \\[16pt]
$^{136}$Xe-$^{136}$Ba & 
3.094 & $-$0.467 & 
$\left\{ \begin{array}{c} 5.040 \\[-2pt]
 3.384 
\end{array}\right.$ & 
$\left\{ \begin{array}{c}1.223 \\[-2pt]
 0.060 
\end{array}\right.$ &
$\left\{\begin{array}{c} 0.49 \\[-2pt]
1.27
\end{array}\right.$
\\[10pt] 
$^{150}$Nd-$^{150}$Sm & 3.380 & $-$0.764 & 
$\left\{\begin{array}{c}6.000 \\[-2pt]
 4.545 \\[-2pt]
 3.854 
\end{array}\right.$ & 
$\left\{ \begin{array}{c} 0.135 \\[-2pt]
 0.047 \\[-2pt]
 0.011 
\end{array}\right.$ & 
$\left\{\begin{array}{c} 0.54 \\[-2pt]
0.81 \\[-2pt]
1.27
\end{array}\right.$
\end{tabular}
\end{ruledtabular}
\end{table}
%
\section{\label{sec:summary} Summary}
In this paper, I investigated the new method to determine the strength of the isoscalar pairing interaction. The principle is the identity relating the double-charge-change and double-pair-transfer ($nn^\prime$ and $pp^\prime$) transitions. This identity is nontrivial to approximations and implies a constraint to the effective interactions used in the approximations. My motivation is to remove an uncertainty in the QRPA calculation of the nuclear matrix element of the $0\nu\beta\beta$ decay. The formulation for this method has been shown in detail including the extension of the QRPA ground state. This new idea has been applied to the nuclei possibly having that decay. It has been demonstrated how the interaction strength can be determined numerically. The GT strength functions were calculated and compared with the experimental data. The $0\nu\beta\beta$ nuclear matrix elements were shown. 

The most important conclusion in the new achievements of this paper is that it is possible to explain the systematics of the strength of the isoscalar pairing interaction determined by my method. The feature is that the strengths of the isoscalar pairing interaction for the magic or near-magic nuclei are relatively weak under the global scaling by $1/A$. The interaction strength reflects on the nuclear structure of the individual nuclei. The next important conclusion is that the interaction strength is not as large as creating the QRPA solutions close to the breaking point in any case investigated. This property is possible because my method refers to the correlations obtained by the lpQRPA. If the HFB ground state is stable against the lpQRPA excitations, it is also stable against the pnQRPA excitations. 
The feature of my method compared to other ones is that this method is applicable to many nuclei consistently with the QRPA. 
The values of the obtained strengths are smaller than those by other method for the $N$$=$$Z$ nuclei in terms of the ratio to the isovector pairing strength. It was assumed for the computation of my method that the protons and neutrons are not in the pair condensate. Thus, it is difficult to discuss whether or not the $N$$=$$Z$ nuclei have the condensate. 
\vfill
\begin{acknowledgments} 
The numerical calculations of this paper were performed by 
the K computer at RIKEN Center for Computational Science, through the program of High Performance Computing Infrastructure (HPCI) in 2016 (hp160052), 2017B (hp170288), 2018B (hp180232).  Computer Oakforest-PACS at Joint Center for Advanced High Performance Computing (JCAHPC) was also used in 2019 (hp190001) through HPCI. Computer Coma at Center for Computational Sciences, University of Tsukuba was used through
Multidisciplinary Cooperative Research Program of this center in 2016 (TKBNDFT) and 2017 (DBTHEORY). 
Furthermore computer Oakforest-PACS at JCAHPC was used through the above program of Center for Computational Sciences, University of Tsukuba in 2018 and 2019 (xg18i006). 
This study is supported by European Regional Development Fund, Project ``Engineering applications of microworld physics" (No. CZ.02.1.01/0.0/0.0/16\_019/0000766) and COREnet of Research Center for Nuclear Physics, Osaka University (project number 8, October 2019$-$March 2020). 
\end{acknowledgments}

\begin{center}
\vspace{4pt}
\textbf{Appendix}
\end{center}
In this appendix, the method is shown to calculate the generalized expectation value of product of the quasiparticle creation and annihilation operators as Eq.~(\ref{eq:example_generalized_exp_value}) using the generalized application of Wick's theorem. It has been proven \cite{Bal69} that 
\begin{eqnarray}
&&\langle f|x_1 x_2 \cdots x_n |i\rangle = \langle f|i\rangle
\sum_{ \substack{ \textrm{\scriptsize{all possible}} \\ \textrm{\scriptsize{contractions}}} }  
\bcontraction{}{x_1}{x_2}{\cdots}
\bcontraction[2ex]{x_1}{x_2}{\cdots}{x_n}
x_1 x_2 \cdots x_n, 
\label{eq:generalized_exp_value_Wick} \\
&&n\textrm{: even integer}, \nonumber
\end{eqnarray}
where $x_1$, $x_2$, and $x_n$ denote any operator of $a^{f\dagger}_\mu$, $a^f_\mu$, $a^{i\dagger}_\mu$, and $a^i_\mu$. The two independent HFB states $|i\rangle$ and $|f\rangle$ are the vacuua of these quasiparticles;
\begin{eqnarray}
a^i_\mu | i\rangle = a^f_\mu | f\rangle = 0.
\end{eqnarray}
The contraction is defined by 
\begin{eqnarray}
\bcontraction{}{x_1}{}{x_2}
x_1 x_2 = \frac{ \langle f|x_1 x_2|i\rangle }{ \langle f|i\rangle },
\end{eqnarray}
with the assumption that $\langle f|i\rangle\neq 0$. 
Equation (\ref{eq:generalized_exp_value_Wick}) holds, if 
\begin{eqnarray}
a^f_\mu |i\rangle \neq 0, \ \langle f|a^{i\dagger}_\mu \neq 0.
\end{eqnarray}

Any product of the fermion operators in which all operators are used for the contractions (I call this product full contraction) can be represented by functions of integers
\begin{eqnarray}
&&\mathcal{I}_n(k;i,j)=1,2,\cdots,n, \\
&&k=1,\cdots, (n-1)!!, \nonumber \\
&&i=1,\cdots,n/2, \nonumber \\
&&j=1,2, \nonumber \\[10pt]
&&\mathcal{P}_n(k)=\pm 1,
\end{eqnarray}
as 
\begin{eqnarray}
\lefteqn{
\bcontraction{}{x_1}{x_2}{x_3}
\bcontraction[2ex]{x_1}{x_2}{x}{\cdots}
\bcontraction{x_1 x_2 x_3}{\cdots}{}{x}
x_1 x_2 x_3 \cdots x_n } 
\nonumber \\
&=&\bcontraction[1.2ex]{}{x\vphantom{g}}{_{\mathcal{I}_n(k;1,1)}}{x}
x_{\mathcal{I}_n(k;1,1)} x_{\mathcal{I}_n(k;1,2)} 
\bcontraction[1.2ex]{}{x}{_{\mathcal{I}_n(k;2,1)}}{x}
x_{\mathcal{I}_n(k;2,1)} x_{\mathcal{I}_n(k;2,2)} \cdots \nonumber \\
&&\times \bcontraction[1.2ex]{}{x}{_{\mathcal{I}_n(k;n/2,1)}}{x}
x_{\mathcal{I}_n(k;n/2,1)} x_{\mathcal{I}_n(k;n/2,2)} \mathcal{P}_n(k), \label{eq:sample_fully_contracted_term}
\end{eqnarray}
where $\mathcal{P}_n(k)$ is the parity of the permutation leading the first line to the second line.  
The first argument $k$ of $\mathcal{I}_n$ specifies a term in the summation
\begin{eqnarray}
\sum_{ \substack{ \textrm{\scriptsize{all possible}} \\ \textrm{\scriptsize{contractions}}} }  
\bcontraction{}{x_1}{x_2}{\cdots}
\bcontraction[2ex]{x_1}{x_2}{\cdots}{x_n}
x_1 x_2 \cdots x_n.  \label{eq:sum_fully_contracted_term}
\end{eqnarray}
The second argument of $\mathcal{I}_n$ indicates the contraction number counted from the leftmost one, and the third argument indicates the left (1) or right (2) operator in the contraction. Namely, $\mathcal{I}_n(k;i,j)$ is the integer corresponding to the quasiparticle state of the operator specified as above and the information of whether the operator is creating or annihilating.  There is a trivial arbitrarity in the definition of $\mathcal{I}_n(k;i,j)$ because of the commutability of the contractions. One of those equivalent ones can be chosen arbitrarily. 

Suppose that $\mathcal{I}_n(k;i,j)$ and $\mathcal{P}_n(k)$ are given. Then, $\mathcal{I}_{n+2}(k;i,j)$ and $\mathcal{P}_{n+2}(k)$ can be constructed using $\mathcal{I}_n(k;i,j)$ and $\mathcal{P}_n(k)$ with extension of the regions of $k$ and $i$ as below. 
Let me consider a product of $n+2$ operators
\begin{eqnarray}
x_1 x_2 \cdots x_{n+2}, \label{eq:operator_product_n+2}
\end{eqnarray}
and introduce one contraction to this operator product;
\begin{eqnarray}
\lefteqn{ \bcontraction{}{x_1}{x_2\cdots}{x_l} 
x_1 x_2 \cdots x_l \cdots x_{n+2} } \nonumber \\
&=&
(-)^{l-2}
\bcontraction{}{x_1}{}{x_l}
x_1 x_l x_2 \cdots x_{l-1} x_{l+1} \cdots x_{n+2}. 
\end{eqnarray}
One can make a full contraction of this product by applying $\mathcal{I}_{n}(k;i,j)$ and $\mathcal{P}_n(k)$ to the product
\begin{eqnarray}
x_2 \cdots x_{l-1} x_{l+1} \cdots x_{n+2},
\end{eqnarray}
with an integer function $\mathcal{J}_l$ :
\begin{eqnarray}
&&\mathcal{J}_l(1) = 2, \cdots, \mathcal{J}_l(l-2) = l-1,\  
\mathcal{J}_l(l-1) = l+1, \cdots, \nonumber \\
&&\mathcal{J}_l(n) = n+2, \nonumber \\
&&(3\leq l \leq n+1), 
\end{eqnarray}
as
\begin{eqnarray}
&&\bcontraction{}{x}{_{\mathcal{J}_l(\mathcal{I}_n(k;1,1))}}{x}
x_{\mathcal{J}_l(\mathcal{I}_n(k;1,1))} x_{\mathcal{J}_l(\mathcal{I}_n(k;1,2))} 
\cdots \nonumber \\
&&\times \bcontraction{}{x}{_{\mathcal{J}_l(\mathcal{I}_n(k;n/2,1))}}{x}
x_{\mathcal{J}_l(\mathcal{I}_n(k;n/2,1))} x_{\mathcal{J}_l(\mathcal{I}_n(k;n/2,2))} \mathcal{P}_n(k). \label{eq:part_product_n+2_contraction}
\end{eqnarray}
For $l=2$ and $n+2$, one can use 
\begin{eqnarray}
&&\mathcal{J}_l(1) = 3, \cdots, \mathcal{J}_l(n) = n+2, \ (l=2), \nonumber \\
&&\mathcal{J}_l(1) = 2, \cdots, \mathcal{J}_l(n) = n+1, \ (l=n+2).  
\end{eqnarray}
[Equation (\ref{eq:part_product_n+2_contraction}) does not need change.]  
It follows that 
\begin{eqnarray}
\lefteqn{ \langle f|x_1 x_2 \cdots x_{n+2}|i\rangle } \nonumber \\
&=& \langle f | i \rangle \sum_l (-)^{l}
\bcontraction{}{x_1}{}{x_l}
x_1 x_l
\sum_k
\bcontraction{}{x}{_{\mathcal{J}_l(\mathcal{I}_n(k;1,1))}}{x}
x_{\mathcal{J}_l(\mathcal{I}_n(k;1,1))} x_{\mathcal{J}_l(\mathcal{I}_n(k;1,2)))}
\cdots \nonumber \\
&&\times\bcontraction{}{x}{_{\mathcal{J}_l(\mathcal{I}_n(k;n/2,1))}}{x}
x_{\mathcal{J}_l(\mathcal{I}_n(k;n/2,1))} x_{\mathcal{J}_l(\mathcal{I}_n(k;n/2,2))}
\mathcal{P}_n(k).
\end{eqnarray}
$\mathcal{I}_{n+2}(k^\prime;i,j)$ and $\mathcal{P}_{n+2}(k^\prime)$ [$\max{k^\prime}=(n+1)!!$] for rewriting this equation to 
\begin{eqnarray}
&&\langle f | i \rangle \sum_{k^\prime}
\bcontraction[1.2ex]{}{x\vphantom{g}}{_{\mathcal{I}_{n+2}(k^\prime;1,1)}}{x}
x_{\mathcal{I}_{n+2}(k^\prime;1,1)} x_{\mathcal{I}_{n+2}(k^\prime;1,2)} 
\cdots \nonumber \\
&&\times \bcontraction[1.2ex]{}{x}{_{\mathcal{I}_{n+2}(k^\prime;n/2+1,1)}}{x}
x_{\mathcal{I}_{n+2}(k^\prime;n/2+1,1)} x_{\mathcal{I}_{n+2}(k^\prime;n/2+1,2)}
\mathcal{P}_{n+2}(k^\prime),  \label{eq:sample_fully_contracted_term_n+2}
\end{eqnarray}
are found to be 
\begin{eqnarray}
&&\mathcal{I}_{n+2}(k^\prime;1,1) = 1, \ 
\mathcal{I}_{n+2}(k^\prime;1,2) = l, \nonumber \\
&&\mathcal{I}_{n+2}(k^\prime;2,1) = \mathcal{J}_l(\mathcal{I}_n(k;1,1)), \nonumber \\ 
&&\mathcal{I}_{n+2}(k^\prime;2,2) = \mathcal{J}_l(\mathcal{I}_n(k;1,2)), \nonumber \\
&&\cdots \nonumber \\
&&\mathcal{I}_{n+2}(k^\prime;n/2+1,1) = \mathcal{J}_l(\mathcal{I}_n(k;n/2,1)), \nonumber \\
&&\mathcal{I}_{n+2}(k^\prime;n/2+1,2) = \mathcal{J}_l(\mathcal{I}_n(k;n/2,2)), \\[10pt]
&&\mathcal{P}_{n+2}(k^\prime) = (-)^{l}\mathcal{P}_n(k), \\
&&(2\leq l \leq n+2, \ n\geq 2). \nonumber
\end{eqnarray}
Integer $k^\prime$ can be assigned to arbitrarily ordered $(k,l)$. 
For $\bcontraction{}{x_1}{}{x_2}x_1 x_2$, one has 
\begin{eqnarray}
&&\mathcal{I}_2(1;1,1)=1, \ \mathcal{I}_2(1;1,2)=2, \nonumber \\
&&\mathcal{P}_2(1) = 1. 
\end{eqnarray} 
Now, any $\mathcal{I}_n(k;i,j)$ and $\mathcal{P}_n(k)$ can be obtained recursively, thus, Eq.~(\ref{eq:generalized_exp_value_Wick}) can be calculated. 
This method is suitable to computation; the program is compact, so that the check is easy, and the extendability to the higher order is high. 

The final step is to calculate the contractions of two operators explicitly. 
I introduce the Bogoliubov transformation from the quasiparticle basis $\{ a^{f\dagger}_\mu, a^f_{-\mu}\}$ to the basis $\{ a^{i\dagger}_\mu, a^i_{-\mu}\}$ 
\begin{eqnarray}
&&a^{i\dagger}_\mu = \sum_{\nu} \left( T^{if1}_{\mu\nu}a^{f\dagger}_{\nu} + T^{if2}_{\mu-\nu}a^f_{-\nu} \right), \nonumber \\
&& a^i_{-\mu} = \sum_{\nu}\left(  T^{if1}_{\mu\nu}t_\mu t^\ast_{\nu}a^f_{-\nu} + T^{if2}_{\mu-\nu}t_\mu t_{-\nu}a^{f\dagger}_{\nu} \right), \label{eq:general_Bogoliubov_tf} \\
&& (j^z_\mu=j^z_{\nu}, \ j^z_\mu>0, \ \pi_\mu=\pi_\nu ), \nonumber
\end{eqnarray}
where $t_\mu$ is a phase defined by the time-reversal operator $\hat{T}$ as 
\begin{eqnarray}
\hat{T} a^\dagger_\mu {\hat{T}}^{-1} = t_\mu a^\dagger_{-\mu}. 
\end{eqnarray}
Once the HFB equations are solved for the nuclei of $|i\rangle$ and $|f\rangle$, $T^{if1}_{\mu\nu}$ and $T^{if2}_{\mu-\nu}$ can be obtained; see, e.g., Ref.~\cite{Rin80}. The efficient method may depend on how the HFB equation is solved. 
The HFB states $|i\rangle$ and $|f\rangle$ have a general relation
\begin{eqnarray}
&&|i\rangle = \frac{1}{\mathcal{N}_i} \exp{\left[  \sum_{\mu\nu}D_{\mu-\nu}a^{f\dagger}_\mu a^{f\dagger}_{-\nu}  \right]}|f\rangle, \label{eq:f2i} \\
&&(j^z_\mu=j^z_\nu, \ j^z_\mu >0, \ \pi_\mu=\pi_\nu), \nonumber \\
&& \mathcal{N}_i = \sqrt{ \det{ (I+D^\dagger D) } }, \label{eq:inv_overlap_fi}
\end{eqnarray} 
where $I$ is the unit matrix. $D$ is the matrix consisting of $D_{\mu -\nu}$ and given by 
\begin{eqnarray}
D=-\left(  \frac{1}{T^{if1}}T^{if2} \right)^\ast, 
\end{eqnarray}
where matrixes $T^{if1}$ and $T^{if2}$ consist of $T^{if1}_{\mu\nu}$ and $T^{if2}_{\mu-\nu}$, respectively. 
Equation (\ref{eq:f2i}) implies
\begin{eqnarray}
\langle f|i\rangle = \frac{1}{\mathcal{N}_i}, 
\end{eqnarray}
thus, $\langle f|i\rangle \neq 0$ is also assumed here.
 The following equations can be used for the overlap calculation discussed in this paper:
\begin{eqnarray}
\langle f | a^f_\mu a^f_{-\nu} | i \rangle = \left\{
\begin{array}{l}
\displaystyle{ -\frac{1}{\mathcal{N}_i}D_{\mu-\nu}, \ (j^z_\mu >0), }\\[13pt]
-\langle f | a^f_{-\nu} a^f_\mu | i \rangle, \ (j^z_\mu < 0), 
\end{array} \right. \label{eq:afaf}
\end{eqnarray}
\begin{eqnarray}
\langle f | a^{i\dagger}_\mu a^{i\dagger}_{-\nu} | i \rangle = \left\{
\begin{array}{l}
\displaystyle{ \frac{1}{\mathcal{N}_i}\sum_{\mu^\prime} T^{if2}_{\mu-\mu^\prime} T^{if1\ast}_{\nu\mu^\prime} t^\ast_\nu t_{\mu^\prime} }\\[13pt]
\displaystyle{ -\sum_{\mu^\prime} T^{if2}_{\mu-\mu^\prime}
  \sum_{\nu^\prime}T^{if2\ast}_{\nu-\nu^\prime}t^\ast_\nu t^\ast_{-\nu^\prime} }\\[13pt]
\ \times\langle f | a^f_{\nu^\prime} a^f_{-\mu^\prime} | i \rangle, \ (j^z_\mu > 0), \\[16pt]
-\langle f | a^{i\dagger}_{-\nu} a^{i\dagger}_\mu | i \rangle, \ (j^z_\mu < 0), 
\end{array}\right. \label{eq:aidaid}
\end{eqnarray}
\begin{eqnarray}
\langle f | a^f_\mu a^{i\dagger}_\nu | i \rangle = \left\{
\begin{array}{l}
\displaystyle{  \frac{1}{\mathcal{N}_i} \Big( T^{if1}_{\nu\mu} -\sum_{\nu^\prime}T^{if2}_{\nu-\nu^\prime}
  D_{\mu-\nu^\prime}\Big), \ (j^z_\mu>0), }\\[20pt]
\displaystyle{ t^\ast_{-\nu} t_{-\mu} \langle f | a^f_{-\mu} a^{i\dagger}_{-\nu} | i \rangle^\ast, \ (j^z_\mu<0).}
 \end{array}\right. \label{eq:afaid}
\end{eqnarray}
\vspace{5pt}
\bibliography{constraint_identity3.3}
\end{document}